\documentclass[journal=jacsat,manuscript=article]{achemso}

\usepackage[version=3]{mhchem} 
\usepackage{xcolor}

\author{Jacob W. Fritsky}
\affiliation{School of Molecular Sciences, Arizona State University, Tempe, Arizona 85287, United States}
\author{Hui-Fei Zhai}
\affiliation{School of Molecular Sciences, Arizona State University, Tempe, Arizona 85287, United States}
\author{Yifeng Zhao}
\affiliation{Department of Physics, Arizona State University, Tempe, Arizona 85287, United States}
\author{Aryan Rauniyar}
\affiliation{Department of Physics, Arizona State University, Tempe, Arizona 85287, United States}
\author{Antia S. Botana}
\affiliation{Department of Physics, Arizona State University, Tempe, Arizona 85287, United States}
\author{Jason F. Khoury}
\affiliation{School of Molecular Sciences, Arizona State University, Tempe, Arizona 85287, United States}
\email{jason.khoury@asu.edu}

\title[]
  {Preferential site ordering alters the magnetic structure of Sm$_3$Ru$_4$Sn$_{13-x}$Ge$_x$ (x = 0 -- 2)} 


\begin{document}


\begin{abstract}

An important aspect of materials research is the ability to tune different physical properties through controlled alloying. The \textit{Ln}$_3$\textit{M}$_4$\textit{X}$_{13}$ (Ln = Lanthanide, M = Transition Metal, X = Tetrel) filled skutterudite family is of interest due to the tunability of its constituent components and their effects on physical properties, such as superconductivity and complex magnetism. In this work, Sm$_3$Ru$_4$Sn$_{13-x}$Ge$_x$ (x = 0 -- 2) was synthesized via excess Sn-flux and characterized using powder and single-crystal X-ray diffraction, magnetometry, X-ray photoelectron spectroscopy, and heat capacity. Sm$_3$Ru$_4$Sn$_{13}$ and its Ge-solid-solution members crystallize in the \emph{Pm}$\bar{3}$\emph{n} space group, which has two unique Wyckoff positions for the tetrel (\textit{X}) site. In the solid solution members, Ge shows preferential occupancy for one of the two Wyckoff sites, reaching $\sim$60$\%$ and 100$\%$ occupancy when x = 1 and 2, respectively. Magnetometry and heat capacity measurements of Sm$_3$Ru$_4$Sn$_{13}$ indicated antiferromagnetic ordering at $T_N$ = 7.3 K. However, Sm$_3$Ru$_4$Sn$_{12}$Ge and Sm$_3$Ru$_4$Sn$_{11}$Ge$_2$ showed notably lower-temperature antiferromagnetic phase transitions with substantial peak-broadening at $T_N$ = 5.5 K and 4.1 K, respectively. These data suggest that alloying Ge into Sm$_3$Ru$_4$Sn$_{13}$ causes magnetic frustration within the structure, likely attributable to a change in the density of states from additional Ge $p$ states at the Fermi level. This work demonstrates that preferentially alloying Ge in Sm$_3$Ru$_4$Sn$_{13-x}$Ge$_x$ allows for more precise tunability of its magnetic structure, elucidating design principles for different quantum phases in intermetallic materials.

\end{abstract}

\section{Introduction}

Structure-property relationships are crucial in solid-state chemistry, demonstrating the link between crystal and electronic structure.\cite{hoffmann1987chemistry} Physical properties in the solid-state are often tuned by creating solid solutions, replacing one atom with another of similar size and charge to modify its lattice parameters. However, controlling where the ``solute'' atom integrates into the structure is often a great challenge in solid-state chemistry, as many solid solutions are disordered. Scientists have developed techniques, either by adjusting synthetic parameters (\emph{e.g.} temperature, cooling time, reagent ratios, etc) or looking to the principles of coordination chemistry, to improve control over how the solute atom integrates into the structure. \cite{kanatzidis2005metal, Phelan2012, Chamorro2018, sunshine1987coordination} Nevertheless, atomically precise control of different coordination environments in the solid-state still poses a great challenge in the field.

Control over which crystallographic site an element will occupy in a solid-state material circumvents several issues with disordered solid solutions. To start, disordered phases often have high defect and vacancy concentrations, making it hard to precisely determine their exact stoichiometry and phase purity.\cite{Larson2023,Simonov2020} High defect and vacancy concentrations also make it challenging to discern the origin of electronic properties in extended solids, such as superconductivity and complex magnetism, because a theoretical model cannot exactly describe their crystal structures.\cite{cao2018unconventional} Developing design principles to synthesize ordered solid solutions, where atoms preferentially occupy different sites, will circumvent these issues through the facile crystal growth of well-behaved crystalline solids with greater control over their physical properties.\cite{Tassanov2024, tassanov2024layered, hodges2018two, hodges2020mixed}

In general, preferential occupancy refers to the selective occupation of a specific crystallographic site within the unit cell by the solute atom in a solid solution. An atom preferentially occupying a crystallographic site can arise due to various factors; for example, KCuZrTe$_2$S utilizes acid-base chemistry to have S occupy a specific crystallographic site when Te occupies multiple sites.\cite{Tassanov2024} In CaBe$_2$Ge$_2$, orbital diffusion stabilizes the structure so that [Be—Ge]$^{-3}$ determines if the CaBe$_2$Ge$_2$ structure is favored over the ThCr$_2$Si$_2$ structure.\cite{Zheng1986} Atomic size also plays a factor in the case of Bi$_2$Te$_{1.6}$S$_{1.4}$, where S displays site preference between the chalcogenide layers. \cite{Ji2012} Using these chemical principles, we can attain atomically precise control of solid-state materials with unique coordination environments and oxidation states, allowing for the fine-tuning of properties such as charge transport, complex magnetism, and photovoltaics.\cite{Witting2020,Pradhan2022, Colin2016}

The \textit{Ln}$_3$\textit{M}$_4$\textit{X}$_{13}$ (Ln = La -- Nd, Sm, Gd; M = Co, Ru, Rh, Os, Ir; X = Sn or Ge) structure has been of interest due to its versatile chemical composition, tunability, and myriad of physical properties, such as superconductivity, heavy fermion behavior, and complex magnetism.\cite{Gumeniuk2018, DominguezMontero2023, nair2018absence, nair2019field, nair2016double, Remeika1980, Oduchi2007, Espinosa1982} In the \textit{Ln}$_3$\textit{M}$_4$Sn$_{13}$ structure, Sn occupies two chemically distinct crystallographic positions, the 2\textit{a} and 24\textit{k} Wyckoff sites. The 2\textit{a} site is located in a Sn(2\textit{a})Sn(24\textit{k})$_{12}$ icosahedron, and the 24\textit{k} site forms a \textit{M}Sn(24\textit{k})$_6$ trigonal prism where Sn(24\textit{k}) is covalently bonded to \textit{M}.\cite{Hodeau1980, Ślebarski2013, Zhong2009} The two unique Wyckoff positions and their bonding environments create a chemical distinction that can be targeted for preferential site occupancy. Preferentially alloying smaller isovalent tetrel atoms (such as Ge) to the 2\textit{a} Wyckoff site can tune the \textit{M} -- Sn(24\textit{k}) bond lengths, thus modifying the electronic structure without changing the overall electron count of the solid. There is already some precedent for preferential occupancy in this structure type, with doping of In on the 24\textit{k} Wyckoff site of Sn in La$_3$Co$_4$Sn$_{13}$.\cite{Neha2016} When incorporating In, the superconducting critical temperature (T$_c$) increased from 2.5 K for La$_3$Co$_4$Sn$_{13}$ to 5.1 K for La$_3$Co$_4$Sn$_{11.7}$In$_{1.3}$.\cite{Neha2016} Due to preferential occupancy, these \textit{M} -- Sn(24\textit{k}) bond lengths are affected more than they otherwise would be if the solid solution was randomly distributed, causing an outsized impact on the bond lengths, and in turn the physical properties.

In the \textit{Ln}$_3$\textit{M}$_4$Sn$_{13}$ structure, the \textit{Ln} atom is within a high symmetry D$_{2d}$ site surrounded by 12 Sn(24\textit{k}) and 4 \textit{M} atoms. The crystal field splitting of the \textit{Ln}$^{3+}$ can be calculated using the Russell--Saunders scheme.\cite{Scheie2021} In the case of Sm$^{3+}$, which has a 4\textit{f}$^5$ electron configuration, the orbital angular momentum (\textit{L} = 5) and the  spin angular momentum (\textit{S} = 5/2) combine for a total angular momentum of J = 5/2. Due to the effect of spin-orbit coupling (SOC) in the structure, there are six degenerate energy states, which split into three doubly degenerate states due to the D$_{2d}$ point symmetry of the Sm$^{3+}$ cation. We performed calculations with the program PyCrystalField to confirm the three doubly degenerate energy states of the Sm$^{3+}$ cation in D$_{2d}$ point symmetry (See \textbf{Table S1} for more information).\cite{Scheie2021} The J-state from this electronic configuration is 1/2, which remains constant in subgroups of D$_{2d}$ (such as C$_{2v}$ and C$_s$), meaning that changes in local symmetry around Sm$^{3+}$ will not affect the electronic state. The constant $J$-state when alloying other atoms on the tetrel sites of Sm$_{3}$\textit{M}$_{4}$\textit{X}$_{13}$ dictates that only chemical pressure will impact the magnetic structure through modifying the \textit{M}--Sn(24\textit{k}) bond lengths.\cite{D1CS00563D} Thus, filled skutterudites with Sm$^{3+}$ cations can act as an initial test case for the development of chemical design principles for preferential occupancy.

Herein, we report the effect of preferential occupancy on the thermal and magnetic properties in the solid-solution series Sm$_{3}$Ru$_{4}$Sn$_{13-x}$Ge$_{x}$ (x = 0 -- 2). We utilized excess Sn-flux to synthesize single crystals of Sm$_{3}$Ru$_{4}$Sn$_{13-x}$Ge$_{x}$, and their chemical composition and morphology were confirmed using scanning electron microscopy (SEM) and electron dispersive spectroscopy (EDS). Powder and single-crystal X-ray diffraction (PXRD and SCXRD) were used to confirm the filled skutterudite structure with preferentially occupied Ge on the tetrel site. SCXRD showed that Ge occupies the 2\textit{a} Wyckoff site with an occupancy of $\sim$60.2 $\%$ and 100 $\%$ for x = 1 and x = 2, respectively. X-ray photoelectron spectroscopy (XPS) data exhibit a shift to higher binding energy as Ge increases, indicating that the Ru--Sn(24\textit{k}) bond lengths are shortening, indirectly modifying the magnetic properties. Magnetometry data indicates that all synthesized members of Sm$_{3}$Ru$_{4}$Sn$_{13-x}$Ge$_{x}$ order antiferromagnetically (AFM) with Curie-Weiss temperatures ($\theta$$_{CW}$) of -122.02 K, -33.35 K, and -42.39 K for x = 0 , 1, and 2, respectively. However, the AFM transition broadens as Ge content increases, likely due to magnetic spin frustration from a quasi-low-dimensional magnetic structure. Heat capacity data (C$_p$) supports the magnetometry data, showing that T$_{N}$ decreases from 7.3 K to 4.1 K with increasing Ge concentration, along with added peak-broadening in x = 1 and 2, likely from spin frustration. Density functional theory (DFT) calculations show an increase in the density of states (DOS) in Sm$_{3}$Ru$_{4}$Sn$_{11}$Ge$_{2}$ compared to Sm$_{3}$Ru$_{4}$Sn$_{13}$ from added Ge $p$ states, suggesting that the purported increase in spin frustration is due to changes in the J--J coupling of Sm$^{3+}$ cations. This work demonstrates how preferential occupancy can be used as a design principle for quantum materials, with changes in chemical pressure as a starting point for investigation.

\section{Experimental Section}

\subsection{Reagents}
All chemicals were used as received from Sigma-Aldrich: samarium powder ($\geq$ 99.7$\%$), ruthenium powder (99.9$\%$), tin shot (99.8$\%$), and germanium chips (99.999$\%$).
\subsection{Synthesis}
Sm$_{3}$Ru$_{4}$Sn$_{13-x}$Ge$_{x}$ was synthesized utilizing excess self-reactive Sn-flux. Sm (0.75 mmol, 0.1128g), Ru (1 mmol, 0.1011g), Sn (20 mmol, 2.3742 g) and Ge (0.1625 mmol, 0.0118 g, for x=1 and 0.325 mmol, 0.236 g, for x = 2) were added to an alumina crucible which was placed in a fused silica tube. The crucible was then covered with base-deactivated fused silica wool to separate the crystals upon centrifuging. Afterward, the tube was back-filled with Ar, vacuumed to $\sim$5 mTorr, and then flame-sealed to create an ampoule. This ampoule was placed in a box furnace and heated to 1000 $^{\circ}$C for 10 h, held at 1000 $^{\circ}$C for 20 h, and then cooled to 600 $^{\circ}$C for 60 h. Upon reaching 600 $^{\circ}$C, the ampoule was centrifuged to separate excess Sn-flux from the Sm$_{3}$Ru$_{4}$Sn$_{13-x}$Ge$_{x}$ crystals. The crystals were sonicated in 6 M HCl for 5--7 h to remove excess Sn flux.
\subsection{Single Crystal X-ray Diffraction}
Single crystals were affixed to a hook using STP Motor Oil and attached to a Bruker Apex-II CCD diffractometer with a Mo K$\alpha$ radiation ($\lambda$ = 0.71073 $\mathring{A}$). Data and unit cell determination were collected using APEX4 Suite software at 100 K. SADABS was used for absorption correction. The structure was solved using ShelXT intrinsic phasing and refined using ShelXL least squares method. The 2\textit{a} Wyckoff site was set to be fully occupied by Ge in Sm$_{3}$Ru$_{4}$Sn$_{11}$Ge$_{2}$.
\subsection{X-ray Photoelectron Spectroscopy (XPS)}
X-ray photoelectron spectroscopy data was collected using a powder sample of Sm$_3$Ru$_4$Sn$_{13-x}$Ge$_x$ on a Kratos Axis Supra+ with an Al K$\alpha$ source. A wide and elemental scan were collected at pass energies of 80 eV and 20 eV, respectively.
\subsection{Density Functional Theory Calculations}
Density functional theory (DFT)-based calculations were performed using the QUANTUM ESPRESSO code\cite{Giannozzi_2009}. The Perdew-Zunger local-density approximation (LDA) exchange-correlation functional was employed \cite{perdew-zunger}. A $6\times6\times6$ $k$-grid density and a plane wave energy cutoff of 60 Ry were used. The self-consistent convergence was set to 10$^{-8}$ Ry. The $4f$ electrons are incorporated in the pseudopotential. The calculations were performed using the experimental structural data for Sm$_3$Ru$_4$Sn$_{13}$. In order to avoid the use of large supercells for  Sm$_3$Ru$_4$Sn$_{11}$Ge$_2$, we used the virtual crystal approximation (VCA) \cite{vca} to simulate Ge doping. A pseudopotential was implemented for the virtual atom $V_{\rm{VCA}}=xV_{\rm{Sn}}+(1-x)V_{\rm{Ge}}$ for $x=0.846$.
\subsection{Magnetometry}
Magnetometry data was collected using the vibrating sample measurement (VSM) option on a Quantum Design Physical Property Measurement (PPMS) System. The Sm$_{3}$Ru$_{4}$Sn$_{13-x}$Ge$_{x}$ crystals were pulverized, placed into a sample holder, and measured between 2 and 300 K.
\subsection{Heat Capacity}
Heat Capacity ($C_p$) data was collected using a Quantum Design Physical Property Measurement System between 2 and 300 K. Apiezon N grease attached the single crystals to the sample stage. Heat capacity under variable magnetic fields was conducted from 0 -- 7 T, with arbitrarily oriented single crystals.

\section{Results and Discussion}

\subsection{Synthesis and Structure}

\begin{figure}[hbt!]
  \includegraphics[width=\textwidth]{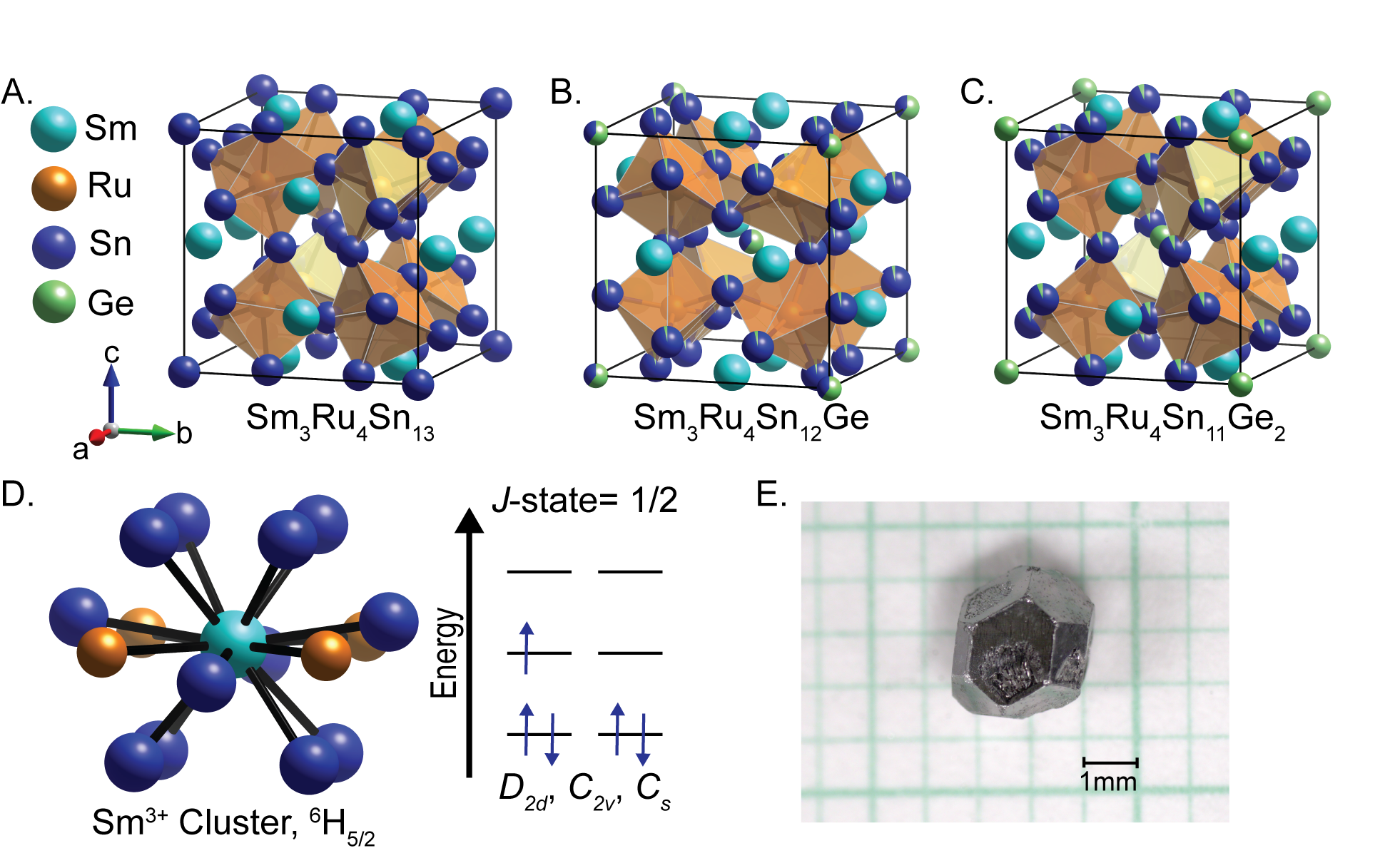}
  \caption{The unit cell of A) Sm$_3$Ru$_4$Sn$_{13}$, B) Sm$_3$Ru$_4$Sn$_{12}$Ge, and C) Sm$_3$Ru$_4$Sn$_{11}$Ge$_2$. All three compounds crystallize in the \emph{Pm}$\bar{3}$\emph{n} space group. Ge preferentially occupies the 2\textit{a} Wyckoff site, filling 60.2\% and 100\% of the site for x = 1 and 2, respectively. D) Coordination environment and $J$-state of the Sm$^{3+}$ ion within the filled skutterudite. Distorting the point group symmetry does not change the electron filling of the $J$-state in Sm$^{3+}$. E) Microscope image of mm-scale as-grown single crystal of Sm$_3$Ru$_4$Sn$_{11}$Ge$_2$.}
  \label{Structure}
  \centering
\end{figure}

Large single crystals ($\sim$3 -- 5 mm) of Sm$_3$Ru$_4$Sn$_{13-x}$Ge$_x$ (x = 0 -- 2) were synthesized via excess Sn-flux, as shown in \textbf{Figure \ref{Structure}}. The composition and morphology of the single crystals were confirmed through scanning electron microscopy (SEM) and energy dispersive spectroscopy (EDS), shown in \textbf{Figure S1}. Attempts to synthesize Sm$_3$Ru$_4$Sn$_{10}$Ge$_3$ were successful but inconsistent, since increasing the concentration of Ge above 2.6 mmol led to recrystallized elemental Ge crashing out of solution. The recrystallized Ge side phase was due to its limited solubility in Sn at the dwell temperature of 1000 $^{\circ}$C, making targeted syntheses of \( x > 2 \) members more difficult.\cite{villars2006asm} Synthesis of high-quality Sm$_3$Ru$_4$Sn$_{13-x}$Ge$_x$ (\( x > 2 \))  single crystals may be possible at temperatures above 1000 $^{\circ}$C, but further investigations of this solid solution family are beyond the scope of this work.

Within Sm$_3$Ru$_4$Sn$_{13}$, there are three primary structural motifs: Sn(2\textit{a})Sn(24\textit{k})$_{12}$ icosahedra, RuSn(24\textit{k})$_{6}$ trigonal prisms, and SmRu$_4$Sn(24\textit{k})$_{12}$ cuboctahedra.\cite{DominguezMontero2023} The Sn(2\textit{a})Sn(24\textit{k})$_{12}$ icosahedra are face sharing with the SmRu$_4$Sn(24\textit{k})$_{12}$ cuboctahedra. In addition, the RuSn(24\textit{k})$_{6}$ trigonal prisms share a face with the icosahedra and cuboctahedra, while also sharing a corner with neighboring trigonal prisms. A motif similar to RuSn(24\textit{k})$_{6}$ can be found in the perovskite (ABX$_3$) structure, where the BX$_6$ octahedra are corner-sharing and surround the 12-coordinate A-site cation.\cite{Hodeau1980} \textbf{{Table \ref{table:Bond Lengths}}} shows selected bond lengths for Sm$_3$Ru$_4$Sn$_{13}$. Within the Sn(2\textit{a})Sn(24\textit{k})$_{12}$ icosahedra, the Sn(2\textit{a}) -- Sn(24\textit{k})$_{12}$ has a bond length of 3.2958(3) \AA. In the RuSn(24\textit{k})$_{6}$ trigonal prism, the Ru -- Sn(24\textit{k}) bond length is 2.63656(12)\AA, which is slightly less than the sum of the covalent radii of 2.65\AA, indicating covalent bonding interactions between Ru -- Sn(24\textit{k}).\cite{Slebarski2015, Mishra2011} The SmRu$_4$Sn(24\textit{k})$_{12}$ bond lengths for Sm -- Sn(24\textit{k}), and Ru -- Sm are 3.3775(3) \AA{} and 3.41158(7) \AA{}, respectively. 

Single-crystal X-ray diffraction confirmed that Sm$_3$Ru$_4$Sn$_{13-x}$Ge$_x$ (x = 0 - 2) crystallizes in the \emph{Pm}$\bar{3}$\emph{n} space group, shown in \textbf{Figure \ref{Structure}A}. Crystallographic data is shown in \textbf{Table \ref{table:Crystal Data}} and \textbf{Table S2}.  The incorporation of Ge into the Sm$_3$Ru$_4$Sn$_{13}$ structure was supported via Rietveld refinements of PXRD patterns, showing a decreasing trend in lattice parameter (\textbf{Figure S2A-C}) from 9.6721(3) \AA{} to 9.6492(4) \AA{} to 9.6347(4) \AA{} for x = 0, 1 and 2, respectively. The main (320) peak (\textbf{Figure S2D}) smoothly transitions to higher angles with increased concentration of Ge, indicating its incorporation into the crystal structure. Ge preferentially occupied the 2\textit{a} Wyckoff site by 60.2$\%$ and 100$\%$ while occupying only 3$\%$ and 6$\%$ of the 24\textit{k} Wyckoff site (shown in \textbf{Figure \ref{Structure}B} and \textbf{Figure \ref{Structure}C}) for x = 1 and 2, respectively. The preference for Ge to occupy the 2\textit{a} Wyckoff site in Sm$_3$Ru$_4$Sn$_{13}$ slightly increases the Sn/Ge(2\textit{a}) -- Sn(24\textit{k}) bond length from 3.2958(3) \AA{} to 3.3030(8) \AA{} to 3.3015(15) \AA (shown in \textbf{Table \ref{table:Bond Lengths}}) for x = 0, 1, and 2, respectively. This increase in bond length is likely attributable to the smaller atomic radius of Ge, which allows for weaker bonding between Ge(2\textit{a}) and Sn(24\textit{k}). In contrast, the bond lengths of Ru -- Sn(24\textit{k}), Sm -- Sn(24\textit{k}), and Ru -- Sm decrease with increased Ge concentration, shown in \textbf{Figure \ref{Structure}D and Table \ref{table:Bond Lengths}}.

\begin{table}[hbt!]
\begin{tabular}{ |c|ccc| } 
\hline 
\textbf{Sample} & \textbf{Sm$_3$Ru$_4$Sn$_{13}$} & \textbf{Sm$_3$Ru$_4$Sn$_{12}$Ge} & \textbf{Sm$_3$Ru$_4$Sn$_{11}$Ge$_2$}\\
\hline
Sn/Ge(2\textit{a}) -- Sn(24\textit{k}) & 3.2958(3) & 3.3030(8)& 3.3015(12)\\ 
Sn(24\textit{k}) -- Sn(24\textit{k}) & 2.9774(6) & 2.9800(14)& 2.975(3)\\ 
Ru -- Sn(24\textit{k}) & 2.63656(12) & 2.6326(3)& 2.6265(5)\\ 
Sm -- Sn(24\textit{k}) & 3.3775(3) & 3.3688(7)& 3.3595(11)\\ 
Ru -- Sm & 3.41158(7) & 3.40486(8)& 3.39588(8)\\
\hline
\end{tabular}
\caption{The bond lengths (\AA) of nearest neighbors X(2\textit{a}) -- Sn(24\textit{k}), Sn(24\textit{k}) -- Sn(24\textit{k}), Ru -- Sn(24\textit{k}),Sm -- Sn(24\textit{k}), Ru -- Sm (X = Sn or Ge) at 100 K.}
\label{table:Bond Lengths}
\end{table}

\begin{table}[hbt!]
\begin{tabular}{ |c|ccc| } 
\hline
\textbf{Empirical Formula} & \textbf{Sm$_3$Ru$_4$Sn$_{13}$} & \textbf{Sm$_3$Ru$_4$Sn$_{12}$Ge} & \textbf{Sm$_3$Ru$_4$Sn$_{11}$Ge$_2$} \\
\hline
Crystal System & \multicolumn{3}{|c|}{Cubic}\\ 
Space Group & \multicolumn{3}{|c|}{\emph{Pm}$\bar{3}$\emph{n}}\\ 
Formula Weight (g/mol) & 2398.3 & 2351.05 & 2317.63\\ 
Calculated Density (g/cm$^3$) & 8.865 & 8.742 & 8.686\\ 
Unit Cell (\AA) & a = 9.6494(2) & a = 9.6304(2) & a = 9.6050(2)\\ 
Volume (\AA$^3$) & 898.46(6) & 893.17(6) & 886.12(6)\\ 
Z & \multicolumn{3}{|c|}{1}\\ 
Temperature (Z) & \multicolumn{3}{|c|}{100}\\
F(000) & 2024 & 1987 & 1961\\
Reflections collected & 12058 & 8327& 8643\\ 
$\theta$ ($^{\circ}$) & 2.985 to 29.971 & 2.991 to 24.983& 2.999 to 24.951\\ 
Absorption coefficient (mm$^{-1}$) & 30.539 & 31.011 & 31.466\\ 
R$_{int}$ & 0.0454 & 0.0402& 0.0506\\ 
Refinement method & \multicolumn{3}{|c|}{Full-matrix least-squares on F$^2$}\\
Final R indices (R$_{obs}$/wR$_{obs}$) & 0.0104/0.0224 & 0.0177/0.0382& 0.0263/0.0554\\ 
R indices (all data) (R$_{all}$/wR$_{all}$) & 0.0110/0.0225 & 0.0177/0.0382& 0.0266/0.0555\\ 
Goodness-of-fit & 1.488 & 1.249& 1.187\\ 
Extinction coefficient & 0.00065(5) & 0.00048(6)& 0.00038(7)\\ 
Largest diff. peak and hole (e$\cdot$\AA$^{-3}$)& 0.751 and -0.519 & 2.053 and -1.720& 2.880 and -3.726\\ 
\hline
\end{tabular}
\caption{Single-crystal data and structural refinement of Sm$_3$Ru$_4$Sn$_{13-x}$Ge$_x$.}
\label{table:Crystal Data}
\end{table}

The electronic ground state with SOC ($J$) of Sm$^{3+}$ for Sm$_3$Ru$_4$Sn$_{13-x}$Ge$_x$ is shown in \textbf{Figure \ref{Structure}D}. The point symmetry of Sm$^{3+}$ is D$_{2d}$ in the \textit{Ln}$_3$\textit{M}$_4$\textit{X}$_{13}$ structure, splitting six degenerate states into three doublets. Since Sm$^{3+}$ has 5 $f$ electrons, the resultant ground state is $J$ = 1/2. Notably, the ground state is also $J$ = 1/2 for subgroups of D$_{2d}$, such as C$_{2v}$ and C$_s$. Given this constant $J$-state, local distortions from the preferential occupancy of Ge will not affect the electron filling of Sm$^{3+}$, making any changes in physical properties a result of chemical pressure from altered bond lengths.  In a sense, this allows for filled skutterudites with Sm$^{3+}$ cations to act as a test case for preferential occupancy: since there is no alteration of the $J$-state from local distortions, we can observe how changes in electronic, thermal, and magnetic properties occur \emph{solely} from chemical pressure. Building off this test case will allow for the development of design principles for preferentially ordered quantum materials in the filled skutterudite structure.

\subsection{XPS}

\begin{figure}[hbt!]
  \includegraphics[width=\textwidth]{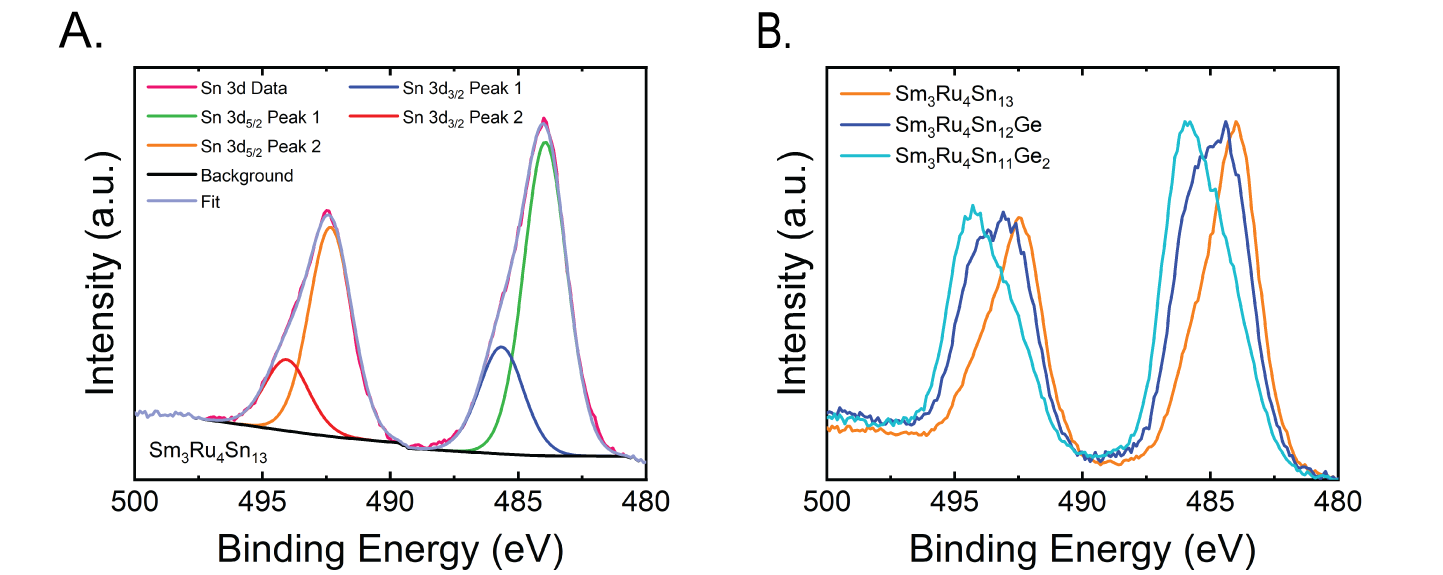}
  \caption{A) X-ray photoelectron spectra (XPS) of Sm$_3$Ru$_4$Sn$_{13}$ for Sn 3\textit{d} and B) Comparison of normalized Sn 3\textit{d} peaks in Sm$_3$Ru$_4$Sn$_{13-x}$Ge$_x$. 
  Increasing the concentration of Ge smoothly changes the contribution from Sn 3\textit{d}$_{5/2}$ to Sn 3\textit{d}$_{3/2}$.}
  \label{XPS}
  \centering
\end{figure}

Now that we have established preferential site ordering of Ge in Sm$_3$Ru$_4$Sn$_{13-x}$Ge$_x$ from X-ray diffraction, we wanted to corroborate our claim with X-ray photoelectron spectroscopy (XPS). The distinct crystallographic positions of Sn and Ge can be distinguished due to the charge density between Ru -- Sn(24\textit{k}) and Sm -- Sn(24\textit{k}).\cite{Slebarski2013,Slebarski2015} \textbf{Figure \ref{XPS}} shows the Sn 3\textit{d} spectra of Sm$_3$Ru$_4$Sn$_{13-x}$Ge$_x$, which shows how the local environment changes with the introduction of Ge. As the Sn(24\textit{k}) atoms shift due to shorter bond lengths between Ru and Sm, the charge density will increase due to stronger covalent bonding. In \textbf{Figure \ref{XPS}A}, the binding energy of Sn 3\textit{d}\textsubscript{3/2} is at $\sim$ 492.5 eV for Sm$_3$Ru$_4$Sn$_{13}$, while Sn 3\textit{d}\textsubscript{5/2} is at $\sim$ 484.2 eV. This is lower than the binding energy for metallic Sn 3\textit{d}$_{5/2}$, which occurs at $\sim$ 485 eV. The 3\textit{d}\textsubscript{5/2} peak is also much larger than the 3\textit{d}\textsubscript{3/2} peak. However, alloying Ge decreases the Sn 3\textit{d}\textsubscript{5/2} peaks while increasing the Sn 3d\textsubscript{3/2} peaks, as shown in \textbf{Figure S3}. There is also a shift to higher binding energies, with Sm$_3$Ru$_4$Sn$_{12}$Ge exhibiting peaks at $\sim$484.2 eV and $\sim$493.1 eV, and Sm$_3$Ru$_4$Sn$_{12}$Ge$_2$ showing peaks at $\sim$485.5 eV and $\sim$494.3 eV for Sn 3\textit{d}\textsubscript{5/2} and Sn 3\textit{d}\textsubscript{3/2}, respectively. This change in peak intensities is most likely due to increased bond strengths between Ru -- Sn(24\textit{k}) and Sm-Sn(24\textit{k}), which is the result of weaker bonding between Ge(2\textit{a}) -- Sn(24\textit{k}). In addition, there is a systematic change in Sn 3d\textsubscript{5/2} and Sn 3d\textsubscript{3/2} from x = 0 to x = 2 (shown in \textbf{Figure \ref{XPS}B}), indicating a smooth transition to stronger covalent bonding between Ru -- Sn(24\textit{k}) and Sm -- Sn(24\textit{k}). 

\subsection{Magnetometry}
\begin{figure}[hbt!]
  \includegraphics[width=0.5\textwidth]{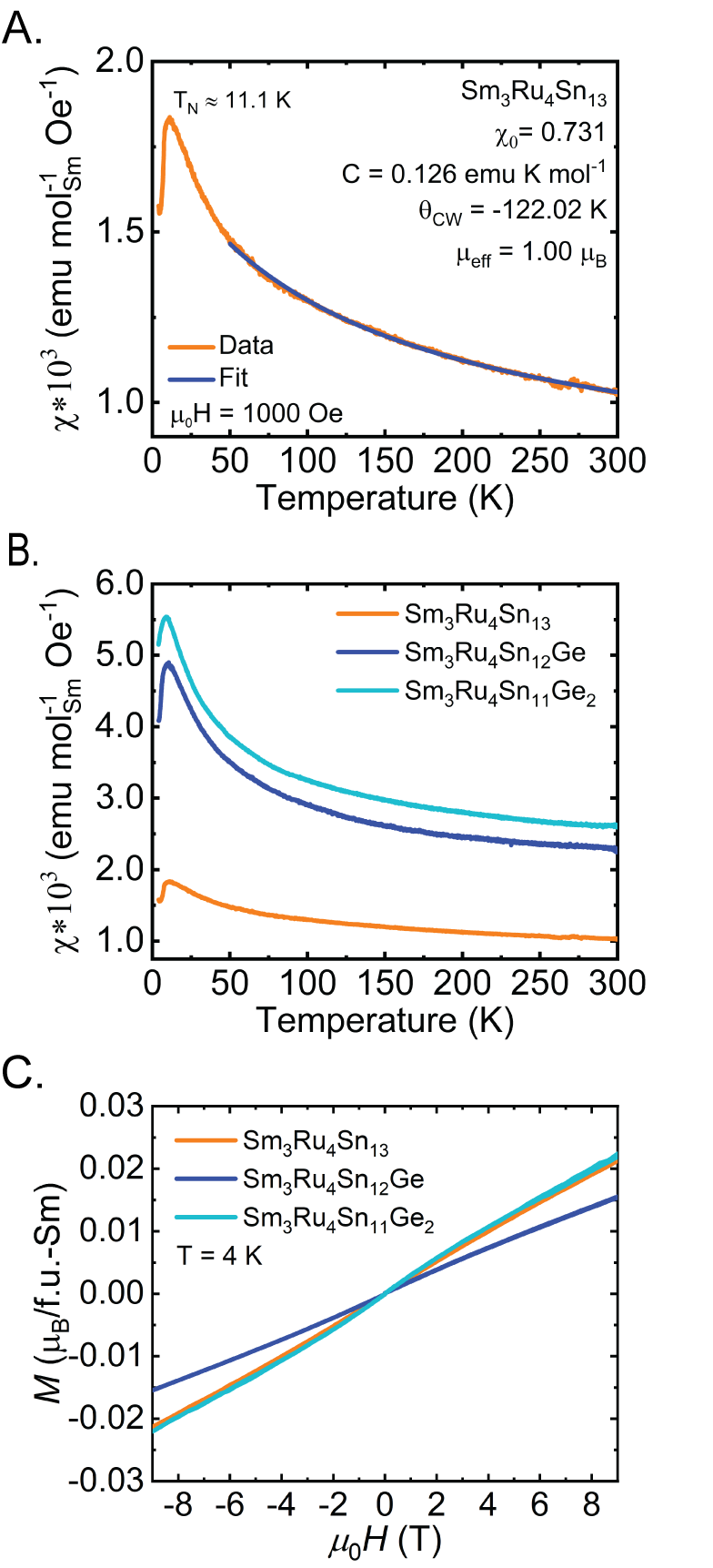}
  \caption{The field-cooled (FC) magnetic susceptibility of A) Sm$_3$Ru$_4$Sn$_{13}$ from 2 -- 300 K. An antiferromagnetic (AFM) transition is present at $\sim$11.1 K. The values from the fitted Curie-Weiss data are $\theta_{CW}$ = -122.02 K and $\mu_{eff}$ = 1.00 $\mu_{B}$. B) A comparison of Sm$_3$Ru$_4$Sn$_{13-x}$Ge$_x$. Each compound shows an AFM transition at $\sim$10.6 K and $\sim$9.5 K for x = 1 and 2, respectively. The $\theta_{CW}$ decreased to -33.35 K and -42.39 K for x = 1 and 2, respectively (Figure S5). C) Magnetic moment ($\mu_{B}$) versus applied magnetic field of Sm$_3$Ru$_4$Sn$_{13-x}$Ge$_x$ from 0 T to 9 T.}
  \label{Magnetometry}
  \centering
\end{figure}

Given how preferential occupancy alters the bonding environment around Sm$^{3+}$, we investigated the magnetic properties of the solid solution. \textbf{Figure \ref{Magnetometry}} shows the magnetic properties arising from the temperature and field-dependent magnetic susceptibility. Field-cooled magnetic susceptibility measurements (FC) and zero-field cooled (ZFC) were identical, so only FC data is shown in \textbf{Figure \ref{Magnetometry}}. The N\'eel temperature (T$_N$) was determined by the onset of the peak in the first derivative plot. In \textbf{Figures \ref{Magnetometry}A} and \textbf{\ref{Magnetometry}B}, a broad antiferromagnetic (AFM) transition is present in all three compounds, with a T$_N$ at approximately $\sim$11.1 K, $\sim$10.6 K, and $\sim$9.3 K for x = 0, 1, and 2, respectively. The broad AFM transitions in Sm$_3$Ru$_4$Sn$_{13-x}$Ge$_x$ are likely the result of local spin frustration, potentially from a quasi-one-dimensional magnetic structure. There is also precedent for these broad, antiferromagnetic peaks in collinear 3D magnets such as Li$_2$CuW$_2$O$_8$, which resemble quasi-1D behavior but have a high degree of magnetic frustration.\cite{ranjith2015collinear} Another possible explanation for the broad AFM transition is spin freezing. Although there are conflicting reports, SmRuSn$_3$ displays spin freezing and shows an antiferromagnetic transition in its resistivity.\cite{Godart1993, Fukuhara_1991} Godart \textit{et. al.} suggest that the conflicting result is due to a difference in lattice parameter, with their reported value of 9.666 \AA {} contrasting with 9.73 \AA{} from Fukuhara \textit{et al.}.\cite{Godart1993, Fukuhara_1991} AC magnetic susceptibility measurements would be needed to confirm the nature of the broad AFM transition Sm$_3$Ru$_4$Sn$_{13-x}$Ge$_x$, as well as the possibility of tuning between spin freezing and antiferromagnetic ordering. \textbf{Figure \ref{Magnetometry}C} shows the magnetization versus applied magnetic field in Sm$_3$Ru$_4$Sn$_{13-x}$Ge$_x$, with no member of the solid solution saturating up to 9 T. The lack of saturated magnetic moments was observed for all three compounds at 4 K, as shown in \textbf{Figure S4}. These robust transitions are consistent with reported magnetic data from Sm$_3$Ru$_4$Ge$_{13}$, along with heat capacity data detailed in the section below.\cite{NAIR2016254} 

Curie-Weiss data for the solid solution was obtained using the following equation:
\[\chi = \frac{C}{T-\theta_{CW}}+\chi_0\]
where C is the Curie constant, T is the absolute temperature, $\theta_{CW}$ is the Curie temperature, and $\chi_0$ is the van Vleck paramagnetism term. Due to Sm$^{3+}$ exhibiting a large temperature-independent van Vleck paramagnetism from the excited J = \(\frac{7}{2}\) multiplet, the Curie-Weiss equation was fitted from 50 K to 300 K with the van Vleck parameter included.\cite{NAIR2016254} Excited state mixing between $^6$H$_{\frac{5}{2}}$ and $^6$H$_{\frac{7}{2}}$ is not expected to contribute to the J-state of the AFM transitions due to their low temperatures, making thermal population less likely. There is also precedent for this claim in SmRuSn$_3$, which has a similar structure and composition to Sm$_3$Ru$_4$Sn$_{13-x}$Ge$_x$ and does not exhibit excited state mixing.\cite{Mazumdar1996TransportSmRuSn_3} From fitting the Curie-Weiss equation (as shown in \textbf{Figure \ref{Magnetometry}A}), we calculated negative Curie-Weiss temperatures ($\theta_{CW}$) for Sm$_3$Ru$_4$Sn$_{13-x}$Ge$_x$, with x = 0, 1, and 2 being -122.02 K, -33.35 K, and -42.39 K, respectively, indicating AFM interactions. The fitted magnet moments ($\mu_{eff}$) are 1.00, 1.04, and 1.12 $\mu_B/mol_{Sm}$ for x = 0, 1, and 2, respectively. (Curie-Weiss fitting for x = 1 and 2 are shown in \textbf{Figure S5}.) Although these fitted magnetic moments are higher than the expected value of 0.85 $\mu_{B}$ for Sm$^{3+}$, they are much closer than the recently reported Sm$_3$Ru$_4$Ge$_{13}$, which has a $\mu_{eff}$ of 0.12 $\mu_{B}$.\cite{NAIR2016254} The magnetic moment of Ru is expected to be quenched, thus not contributing to the higher $\mu_{eff}$ observed.\cite{Mugiraneza2022} The high $\mu_{eff}$ may also be attributable to crystal electric field effects, which play a role in other Sm$^{3+}$--containing compounds such as SmRu$_2$Al$_{10}$ and Sm$_3$ZrBi$_5$.\cite{Peratheepan_2015, khoury2022class}

\subsection{Heat Capacity}
\begin{figure}[hbt!]
  \includegraphics[width=0.9\textwidth]{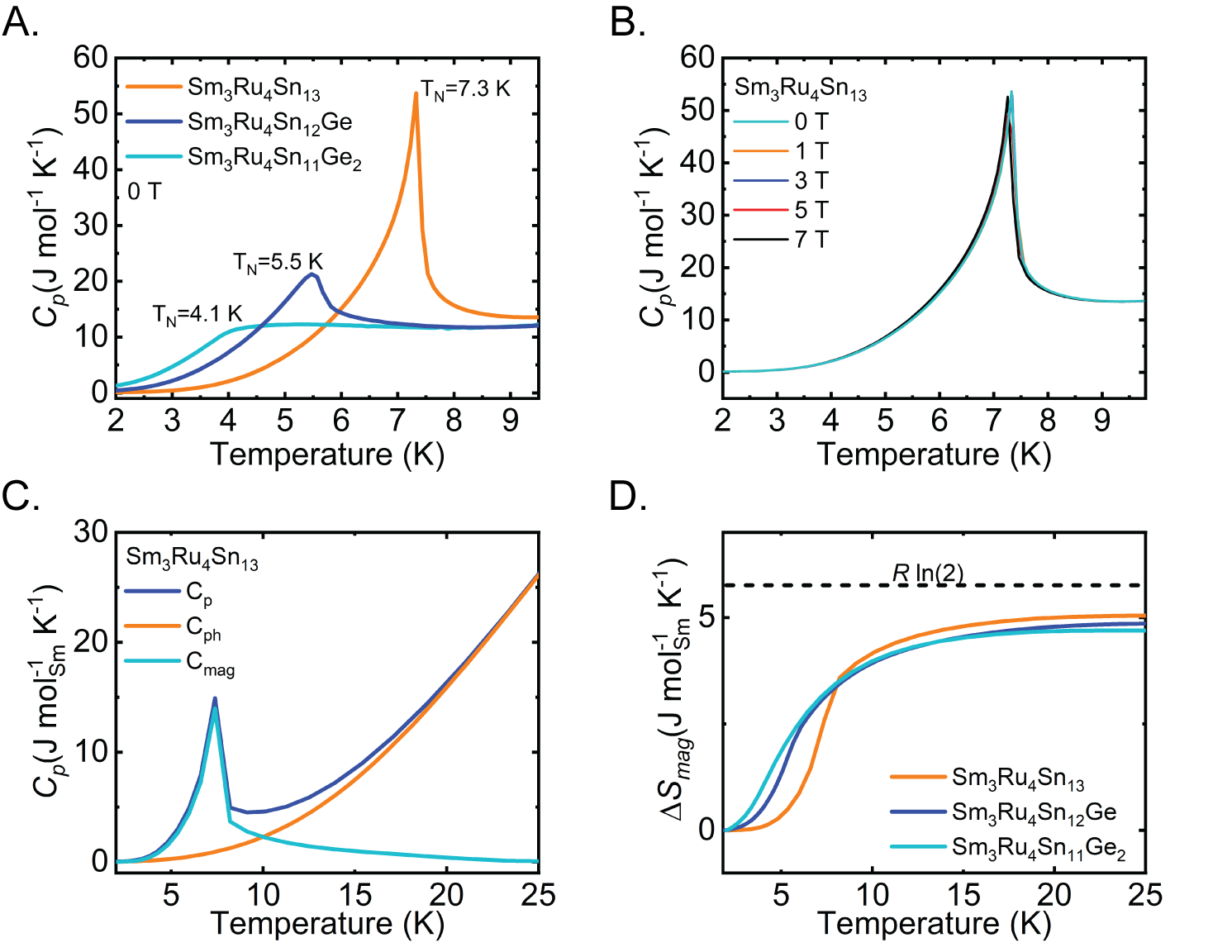}
  \caption{Heat capacity for Sm$_3$Ru$_4$Sn$_{13-x}$Ge$_x$. A) The N\'eel temperature (at 0 T) transitions from T$_N$ = 7.3 K to 4.1 K as the concentration of Ge increases. Peak broadening is also present, which suggests Ge incorporation causes magnetic frustration. B) Sm$_3$Ru$_4$Sn$_{13}$ exhibits field-insensitive behavior up to 7 T. C) The fittings for C$_{ph}$ and C$_{mag}$ using a two-mode  Debye fit on Sm$_3$Ru$_4$Sn$_{13}$. The Debye temperatures are approximately 313 K and 143 K, and the sum of the oscillator terms is $\sim$20.4, in good agreement with the stoichiometry of Sm$_3$Ru$_4$Sn$_{13}$. D) The magnetic entropy (S$_{mag}$) for all three compounds. Each compound approaches \textit{R}ln(2) (5.76 J mol$^{-1}$ K$^{-1}$), indicating $J$ = \(\frac{1}{2}\), consistent with the predicted value for Sm$^{3+}$.}
  \label{Heat Capacity}
  \centering
\end{figure}

Due to the broad antiferromagnetic peaks in the magnetometry data, heat capacity experiments were needed to provide further insight into the magnetic character of these materials. \textbf{Figure \ref{Heat Capacity}A} shows the N\'eel temperatures (T$_N$) of Sm$_3$Ru$_4$Sn$_{13-x}$Ge$_x$ at T$_N$ = 7.3 K, 5.5 K, and 4.1 K for x = 0, 1, and 2, respectively. These antiferromagnetic peaks provide greater accuracy of the continuous phase transition temperatures compared to the magnetic susceptibility data. The sharp $\lambda$-like peak at 7.3 K indicates that long-range magnetic ordering is present in Sm$_3$Ru$_4$Sn$_{13}$. The peak broadening shown in the x = 1 and 2 members of Sm$_3$Ru$_4$Sn$_{13-x}$Ge$_x$ suggests that alloying Ge causes magnetic frustration even though there is preferential site ordering, making site-occupancy disorder an unlikely explanation for this result. Surprisingly, variable-field data taken from 0 -- 7 T (\textbf{Figure \ref{Heat Capacity}B}) show that the transition in Sm$_3$Ru$_4$Sn$_{13}$ is field-insensitive, likely attributable to crystal electric field effects from the local symmetry of the Sm$^{3+}$ cation.\cite{nair2019field} This invariance to magnetic field is consistent in the x = 1 and 2 members of Sm$_3$Ru$_4$Sn$_{13-x}$Ge$_x$ (\textbf{Figure S6}), along with other Sm-containing compounds such as Sm$_3$BiZr$_5$, SmTi$_2$Al$_{20}$, and SmOs$_4$Sb$_{12}$, lending credence to the local symmetry being preserved in the preferentially occupied compounds.\cite{khoury2022class, Higashinaka2011, sanada2005exotic} Surprisingly, Sm$_3$Ru$_4$Sn$_{11}$Ge$_2$ has a lower T$_N$ than the fully Ge analog, Sm$_3$Ru$_4$Ge$_{13}$, which has a sharp AFM peak at T$_N$ = 5 K.\cite{NAIR2016254} Therefore, tuning the magnetic properties in the Sm$_3$Ru$_4$Sn$_{13}$ system does not follow Vegard's law between Sn and Ge, possibly allowing for preferential occupancy to modify the magnetic behavior more than it otherwise would in a perfectly disordered solid solution.\cite{denton1991vegard} Further studies would be needed to determine the evolution of the magnetic structure as Sm$_3$Ru$_4$Sn$_{13-x}$Ge$_x$ approaches full occupation of Ge.

The value of the contribution of the phonon to the heat capacity (C$_{ph}$) of Sm$_3$Ru$_4$Sn$_{13-x}$Ge$_x$ was calculated by fitting a two-mode Debye fit (\textbf{Figure S6}) from 25 K to 150 K, as shown below:
\[C_{ph} = 9s_1R(\frac{T}{\theta_{D1}})^3\int_{0}^{\frac{\theta_{D1}}{T}}\frac{x^4e^x}{(e^x-1)^2}dx+9s_2R(\frac{T}{\theta_{D2}})^3\int_{0}^{\frac{\theta_{D2}}{T}}\frac{x^4e^x}{(e^x-1)^2}dx\]
where \textit{R} is the ideal gas constant, $T$ is the temperature, $s$ is the oscillator strength, and $\theta_D$ is the Debye temperature. A two-mode Debye fit was selected because a single-mode Debye fit did not adequately fit the data, suggesting a more complex phonon background in Sm$_3$Ru$_4$Sn$_{13-x}$Ge$_x$. The sum of the two oscillator terms ($s_1$ and $s_2$) is $\sim$20 in all three compounds, in good agreement with the stoichiometry Sm$_3$Ru$_4$Sn$_{13-x}$Ge$_x$. From the calculation of C$_{ph}$, the magnetic contribution to the heat capacity (C$_{mag}$) can be calculated by subtracting C$_{ph}$ from the total of C$_p$. Then, the magnetic entropy (S$_{mag}$) can be calculated by integrating C$_{mag}$ (\(\frac{C_{mag}}{T})\) over the temperature range of the magnetic transition to determine the $J$-state of Sm$^{3+}$ in the filled skutterudite structure. C$_{ph}$ and C$_{mag}$ are shown in \textbf{Figure \ref{Heat Capacity}C} with $\theta_{D1}$ and $\theta_{D2}$ equal to 313 K and 143 K, respectively, for x = 0. These Debye temperatures have similar values in the x = 1 and 2 members of the solid solution (\textbf{Figure S6}), indicating similar phonon modes.  The magnetic entropy (S$_{mag}$) for all three compounds (shown in \textbf{Figure \ref{Heat Capacity}D}) saturate near \textit{R}ln(2) (5.76 J mol$^{-1}$ K$^{-1}$), which is consistent with a $J$ = \(\frac{1}{2}\) for Sm$^{3+}$, supporting that all changes in the magnetic structure of Sm$_3$Ru$_4$Sn$_{13-x}$Ge$_x$ are due to chemical pressure from incorporation of Ge.

\subsection{Density Functional Theory Calculations}
\begin{figure}[hbt!]
  \includegraphics[width=0.5\textwidth]{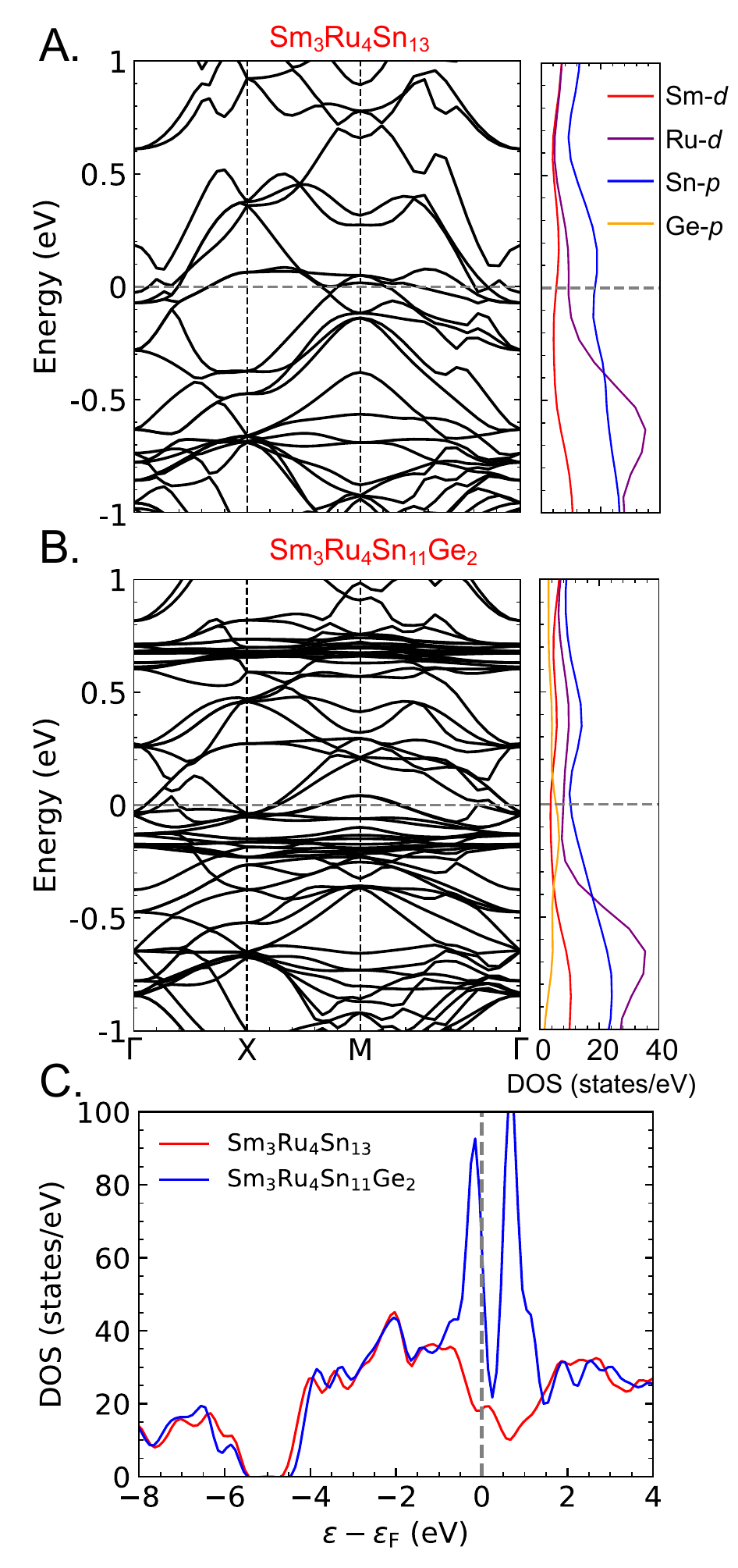}
  \caption{Electronic band structure and density of states (DOS) for (A) Sm$_3$Ru$_4$Sn$_{13}$ and (B) Sm$_3$Ru$_4$Sn$_{11}$Ge$_2$. The right panels in each figure correspond to the orbital-resolved density of states. (C) The total density states of Sm$_3$Ru$_4$Sn$_{13}$ and Sm$_3$Ru$_4$Sn$_{11}$Ge$_2$ compounds. The Fermi energy has been set to zero.}
  \label{DFT}
  \centering
\end{figure}

\textbf{Figure \ref{DFT}} shows the calculated band structure and atom-resolved density of states (DOS) for Sm$_3$Ru$_4$Sn$_{13}$ (\textbf{Figure \ref{DFT}A}) and Sm$_3$Ru$_4$Sn$_{11}$Ge$_2$ (\textbf{Figure \ref{DFT}B}) in the paramagnetic state. The most noticeable modification upon Ge incorporation is a drastic change in the DOS (\textbf{Figure \ref{DFT}C}) at the Fermi level ($\epsilon_F$) and the concomitant addition of extra bands of Ge-$p$ character in the vicinity of $\epsilon_F$. While the role and interaction of the $4f$ electrons of Sm with the other valence electrons cannot be discussed within the framework of the DFT calculations we used (these are paramagnetic calculation and, as mentioned in the Methods, the $4f$ electrons of Sm are incorporated in the pseudopotential), we argue that such a large change in the DOS might affect the $J-J$ coupling of the Sm$^{3+}$ cations  and could be at the origin of the broadening and shift in the magnetic transition described above.

\section{Conclusion}
In conclusion, we determined that preferential occupancy has an outsized effect on the magnetic structure of Sm$_3$Ru$_4$Sn$_{13-x}$Ge$_x$, with Ge selectively occupying one of two tetrel crystallographic sites. Structural analysis showed that Ge preferred the 2\textit{a} Wyckoff site by $\sim$60.2$\%$ and 100$\%$ for x = 1 and 2, respectively, with the occupancy of the 24\textit{k} site only $\sim$3$\%$ and $\sim$6$\%$ Ge in those two compounds. XPS data exhibited a smooth transition to a higher binding energy, which can be attributed to an increased bond strength between Ru -- Sn(24\textit{k}). Broad AFM transitions were present in the magnetometry data for all three compounds, suggesting a quasi-1D magnetic structure of Sm$^{3+}$ cations. Heat capacity measurements showed dramatic changes in magnetic ordering temperature and peak broadening as a function of preferentially occupied Ge. The AFM transitions were invariant to magnetic field up to 7 T, likely due to crystal electric field effects of the local symmetry around the Sm$^{3+}$ cation. Despite the ordered nature of these solid solutions, spin frustration may be attributable to an increase in the DOS from additional Ge $p$ states, changing the J--J coupling of the RKKY interactions. Further studies on structure-property relationships in the \textit{Ln}$_3$\textit{M}$_4$Sn$_{13}$ family and its solid solutions, such as changes to the tetrel and lanthanide sites, are needed to fully outline design principles for constructing site-ordered magnetic quantum materials.

\begin{acknowledgement}

J.F.K. would like to thank the School of Molecular Sciences at Arizona State University for providing start-up funding for this research. We acknowledge Marco Flores for assisting with the preparation of single crystal samples for single crystal X-ray diffraction. We also acknowledge the use of facilities within the Eyring Materials Center at Arizona State University. A.S.B. was supported by the Alfred P. Sloan Foundation (FG-2022-19086).

\end{acknowledgement}

\begin{suppinfo}

The Supporting Information is available free of charge. The Supporting Information includes additional PXRD, XPS, magnetometry, and heat capacity data.

\end{suppinfo}


\begin{mcitethebibliography}{50}
\providecommand*\natexlab[1]{#1}
\providecommand*\mciteSetBstSublistMode[1]{}
\providecommand*\mciteSetBstMaxWidthForm[2]{}
\providecommand*\mciteBstWouldAddEndPuncttrue
  {\def\EndOfBibitem{\unskip.}}
\providecommand*\mciteBstWouldAddEndPunctfalse
  {\let\EndOfBibitem\relax}
\providecommand*\mciteSetBstMidEndSepPunct[3]{}
\providecommand*\mciteSetBstSublistLabelBeginEnd[3]{}
\providecommand*\EndOfBibitem{}
\mciteSetBstSublistMode{f}
\mciteSetBstMaxWidthForm{subitem}{(\alph{mcitesubitemcount})}
\mciteSetBstSublistLabelBeginEnd
  {\mcitemaxwidthsubitemform\space}
  {\relax}
  {\relax}

\bibitem[Hoffmann(1987)]{hoffmann1987chemistry}
Hoffmann,~R. How chemistry and physics meet in the solid state. \emph{Angew. Chem. Int. Ed.} \textbf{1987}, \emph{26}, 846--878\relax
\mciteBstWouldAddEndPuncttrue
\mciteSetBstMidEndSepPunct{\mcitedefaultmidpunct}
{\mcitedefaultendpunct}{\mcitedefaultseppunct}\relax
\EndOfBibitem
\bibitem[Kanatzidis \latin{et~al.}(2005)Kanatzidis, P{\"o}ttgen, and Jeitschko]{kanatzidis2005metal}
Kanatzidis,~M.~G.; P{\"o}ttgen,~R.; Jeitschko,~W. The metal flux: a preparative tool for the exploration of intermetallic compounds. \emph{Angew. Chem. Int. Ed.} \textbf{2005}, \emph{44}, 6996--7023\relax
\mciteBstWouldAddEndPuncttrue
\mciteSetBstMidEndSepPunct{\mcitedefaultmidpunct}
{\mcitedefaultendpunct}{\mcitedefaultseppunct}\relax
\EndOfBibitem
\bibitem[Phelan \latin{et~al.}(2012)Phelan, Menard, Kangas, McCandless, Drake, and Chan]{Phelan2012}
Phelan,~W.~A.; Menard,~M.~C.; Kangas,~M.~J.; McCandless,~G.~T.; Drake,~B.~L.; Chan,~J.~Y. Adventures in crystal growth: Synthesis and characterization of single crystals of complex intermetallic compounds. \emph{Chem. Mater.} \textbf{2012}, \emph{24}, 409--420\relax
\mciteBstWouldAddEndPuncttrue
\mciteSetBstMidEndSepPunct{\mcitedefaultmidpunct}
{\mcitedefaultendpunct}{\mcitedefaultseppunct}\relax
\EndOfBibitem
\bibitem[Chamorro and McQueen(2018)Chamorro, and McQueen]{Chamorro2018}
Chamorro,~J.~R.; McQueen,~T.~M. Progress toward Solid State Synthesis by Design. \emph{Acc. Chem. Res.} \textbf{2018}, \emph{51}, 2918--2925\relax
\mciteBstWouldAddEndPuncttrue
\mciteSetBstMidEndSepPunct{\mcitedefaultmidpunct}
{\mcitedefaultendpunct}{\mcitedefaultseppunct}\relax
\EndOfBibitem
\bibitem[Sunshine \latin{et~al.}(1987)Sunshine, Keszler, and Ibers]{sunshine1987coordination}
Sunshine,~S.~A.; Keszler,~D.~A.; Ibers,~J.~A. Coordination chemistry and the solid state. \emph{Acc. Chem. Res.} \textbf{1987}, \emph{20}, 395--400\relax
\mciteBstWouldAddEndPuncttrue
\mciteSetBstMidEndSepPunct{\mcitedefaultmidpunct}
{\mcitedefaultendpunct}{\mcitedefaultseppunct}\relax
\EndOfBibitem
\bibitem[Larson and Latturner(2023)Larson, and Latturner]{Larson2023}
Larson,~J.~T.; Latturner,~S.~E. Flux Growth of an Intermetallic with Interstitial Fluorides via Decomposition of a Fluorocarbon. \emph{Inorg. Chem.} \textbf{2023}, \emph{62}, 1508--1512, PMID: 36634226\relax
\mciteBstWouldAddEndPuncttrue
\mciteSetBstMidEndSepPunct{\mcitedefaultmidpunct}
{\mcitedefaultendpunct}{\mcitedefaultseppunct}\relax
\EndOfBibitem
\bibitem[Simonov and Goodwin(2020)Simonov, and Goodwin]{Simonov2020}
Simonov,~A.; Goodwin,~A.~L. Designing disorder into crystalline materials. \emph{Nat. Rev. Chem.} \textbf{2020}, \emph{4}, 657--673\relax
\mciteBstWouldAddEndPuncttrue
\mciteSetBstMidEndSepPunct{\mcitedefaultmidpunct}
{\mcitedefaultendpunct}{\mcitedefaultseppunct}\relax
\EndOfBibitem
\bibitem[Cao \latin{et~al.}(2018)Cao, Fatemi, Fang, Watanabe, Taniguchi, Kaxiras, and Jarillo-Herrero]{cao2018unconventional}
Cao,~Y.; Fatemi,~V.; Fang,~S.; Watanabe,~K.; Taniguchi,~T.; Kaxiras,~E.; Jarillo-Herrero,~P. Unconventional superconductivity in magic-angle graphene superlattices. \emph{Nature} \textbf{2018}, \emph{556}, 43--50\relax
\mciteBstWouldAddEndPuncttrue
\mciteSetBstMidEndSepPunct{\mcitedefaultmidpunct}
{\mcitedefaultendpunct}{\mcitedefaultseppunct}\relax
\EndOfBibitem
\bibitem[Tassanov \latin{et~al.}(2024)Tassanov, Lee, Xia, and Hodges]{Tassanov2024}
Tassanov,~A.; Lee,~H.; Xia,~Y.; Hodges,~J.~M. Rational Pathways to Ordered Multianion Chalcogenides Using Retrosynthetic Crystal Chemistry. \emph{J. Am. Chem. Soc.} \textbf{2024}, \emph{146}, 32627--32639\relax
\mciteBstWouldAddEndPuncttrue
\mciteSetBstMidEndSepPunct{\mcitedefaultmidpunct}
{\mcitedefaultendpunct}{\mcitedefaultseppunct}\relax
\EndOfBibitem
\bibitem[Tassanov \latin{et~al.}(2024)Tassanov, Lee, Xia, and Hodges]{tassanov2024layered}
Tassanov,~A.; Lee,~H.; Xia,~Y.; Hodges,~J.~M. Layered NaBa$_2$M$_3$Q$_3$ (Q$_2$)(M= Cu or Ag; Q= S or Se) Chalcogenides and Local Ordering in Their Mixed-Anion Compositions. \emph{Inorg. Chem.} \textbf{2024}, \emph{63}, 15584--15591\relax
\mciteBstWouldAddEndPuncttrue
\mciteSetBstMidEndSepPunct{\mcitedefaultmidpunct}
{\mcitedefaultendpunct}{\mcitedefaultseppunct}\relax
\EndOfBibitem
\bibitem[Hodges \latin{et~al.}(2018)Hodges, Xia, Malliakas, Alexander, Chan, and Kanatzidis]{hodges2018two}
Hodges,~J.~M.; Xia,~Y.; Malliakas,~C.~D.; Alexander,~G.~C.; Chan,~M.~K.; Kanatzidis,~M.~G. Two-Dimensional CsAg$_5$Te$_{3-x}$S$_x$ Semiconductors: Multi-anion Chalcogenides with Dynamic Disorder and Ultralow Thermal Conductivity. \emph{Chem. Mater.} \textbf{2018}, \emph{30}, 7245--7254\relax
\mciteBstWouldAddEndPuncttrue
\mciteSetBstMidEndSepPunct{\mcitedefaultmidpunct}
{\mcitedefaultendpunct}{\mcitedefaultseppunct}\relax
\EndOfBibitem
\bibitem[Hodges \latin{et~al.}(2020)Hodges, Xia, Malliakas, Slade, Wolverton, and Kanatzidis]{hodges2020mixed}
Hodges,~J.~M.; Xia,~Y.; Malliakas,~C.~D.; Slade,~T.~J.; Wolverton,~C.; Kanatzidis,~M.~G. Mixed-valent copper chalcogenides: tuning structures and electronic properties using multiple anions. \emph{Chem. Mat.} \textbf{2020}, \emph{32}, 10146--10154\relax
\mciteBstWouldAddEndPuncttrue
\mciteSetBstMidEndSepPunct{\mcitedefaultmidpunct}
{\mcitedefaultendpunct}{\mcitedefaultseppunct}\relax
\EndOfBibitem
\bibitem[Zheng and Hoffmann(1986)Zheng, and Hoffmann]{Zheng1986}
Zheng,~C.; Hoffmann,~R. Donor-Acceptor Layer Formation and Lattice Site Preference in the Solid: The CaBe$_2$Ge$_2$ Structure. \emph{J. Am. Chem. Soc} \textbf{1986}, \emph{108}, 3078--3088\relax
\mciteBstWouldAddEndPuncttrue
\mciteSetBstMidEndSepPunct{\mcitedefaultmidpunct}
{\mcitedefaultendpunct}{\mcitedefaultseppunct}\relax
\EndOfBibitem
\bibitem[Ji \latin{et~al.}(2012)Ji, Allred, Fuccillo, Charles, Neupane, Wray, Hasan, and Cava]{Ji2012}
Ji,~H.; Allred,~J.~M.; Fuccillo,~M.~K.; Charles,~M.~E.; Neupane,~M.; Wray,~L.~A.; Hasan,~M.~Z.; Cava,~R.~J. Bi$_2$Te$_{1.6}$S$_{1.4}$: A topological insulator in the tetradymite family. \emph{Phys. Rev. B} \textbf{2012}, \emph{85}, 201103\relax
\mciteBstWouldAddEndPuncttrue
\mciteSetBstMidEndSepPunct{\mcitedefaultmidpunct}
{\mcitedefaultendpunct}{\mcitedefaultseppunct}\relax
\EndOfBibitem
\bibitem[Witting \latin{et~al.}(2020)Witting, Ricci, Chasapis, Hautier, and Snyder]{Witting2020}
Witting,~I.~T.; Ricci,~F.; Chasapis,~T.~C.; Hautier,~G.; Snyder,~G.~J. The Thermoelectric Properties of n-Type Bismuth Telluride: Bismuth Selenide Alloys Bi$_2$Te$_{3-x}$Se$_x$. \emph{Research} \textbf{2020}, \emph{2020}\relax
\mciteBstWouldAddEndPuncttrue
\mciteSetBstMidEndSepPunct{\mcitedefaultmidpunct}
{\mcitedefaultendpunct}{\mcitedefaultseppunct}\relax
\EndOfBibitem
\bibitem[Pradhan \latin{et~al.}(2022)Pradhan, Jena, and Samal]{Pradhan2022}
Pradhan,~A.; Jena,~M.~K.; Samal,~S.~L. Understanding of the Band Gap Transition in Cs$_3$Sb$_2$Cl$_{9-x}$Br$_x$: Anion Site Preference-Induced Structural Distortion. \emph{ACS Appl. Energy Mater.} \textbf{2022}, \emph{5}, 6952--6961\relax
\mciteBstWouldAddEndPuncttrue
\mciteSetBstMidEndSepPunct{\mcitedefaultmidpunct}
{\mcitedefaultendpunct}{\mcitedefaultseppunct}\relax
\EndOfBibitem
\bibitem[Colin \latin{et~al.}(2016)Colin, Ito, Yano, Dempsey, Suard, and Givord]{Colin2016}
Colin,~C.~V.; Ito,~M.; Yano,~M.; Dempsey,~N.~M.; Suard,~E.; Givord,~D. Solid-solution stability and preferential site-occupancy in (R-R\textquotesingle)$_2$Fe$_{14}$B compounds. \emph{Appl. Phys. Lett.} \textbf{2016}, \emph{108}, 43\relax
\mciteBstWouldAddEndPuncttrue
\mciteSetBstMidEndSepPunct{\mcitedefaultmidpunct}
{\mcitedefaultendpunct}{\mcitedefaultseppunct}\relax
\EndOfBibitem
\bibitem[Gumeniuk(2018)]{Gumeniuk2018}
Gumeniuk,~R. Structural and Physical Properties of Remeika Phases. \emph{Handbook on the Physics and Chemistry of Rare Earths} \textbf{2018}, \emph{54}, 43--143\relax
\mciteBstWouldAddEndPuncttrue
\mciteSetBstMidEndSepPunct{\mcitedefaultmidpunct}
{\mcitedefaultendpunct}{\mcitedefaultseppunct}\relax
\EndOfBibitem
\bibitem[Montero \latin{et~al.}(2023)Montero, McCandless, Oladehin, Baumbach, and Chan]{DominguezMontero2023}
Montero,~A.~D.; McCandless,~G.~T.; Oladehin,~O.; Baumbach,~R.~E.; Chan,~J.~Y. Development of a Geometric Descriptor for the Strategic Synthesis of Remeika Phases. \emph{Chem. Mater.} \textbf{2023}, \emph{35}, 2238--2247\relax
\mciteBstWouldAddEndPuncttrue
\mciteSetBstMidEndSepPunct{\mcitedefaultmidpunct}
{\mcitedefaultendpunct}{\mcitedefaultseppunct}\relax
\EndOfBibitem
\bibitem[Nair \latin{et~al.}(2018)Nair, Ogunbunmi, Ghosh, Adroja, Koza, Guidi, and Strydom]{nair2018absence}
Nair,~H.~S.; Ogunbunmi,~M.~O.; Ghosh,~S.~K.; Adroja,~D.~T.; Koza,~M.~M.; Guidi,~T.; Strydom,~A.~M. Absence of a long-range ordered magnetic ground state in Pr$_3$Rh$_4$Sn$_{13}$ studied through specific heat and inelastic neutron scattering. \emph{J. Phys. Condens. Matter} \textbf{2018}, \emph{30}, 145601\relax
\mciteBstWouldAddEndPuncttrue
\mciteSetBstMidEndSepPunct{\mcitedefaultmidpunct}
{\mcitedefaultendpunct}{\mcitedefaultseppunct}\relax
\EndOfBibitem
\bibitem[Nair \latin{et~al.}(2019)Nair, Kumar, Sahu, Xhakaza, Mishra, Samal, Ghosh, Sekhar, and Strydom]{nair2019field}
Nair,~H.~S.; Kumar,~K.~R.; Sahu,~B.; Xhakaza,~S.~P.; Mishra,~P.; Samal,~D.; Ghosh,~S.~K.; Sekhar,~B.~R.; Strydom,~A.~M. Field-Independent Features in the Magnetization and Specific Heat of Sm$_3$Co$_4$Ge$_{13}$. \emph{Crystals} \textbf{2019}, \emph{9}, 322\relax
\mciteBstWouldAddEndPuncttrue
\mciteSetBstMidEndSepPunct{\mcitedefaultmidpunct}
{\mcitedefaultendpunct}{\mcitedefaultseppunct}\relax
\EndOfBibitem
\bibitem[Nair \latin{et~al.}(2016)Nair, Ghosh, and Strydom]{nair2016double}
Nair,~H.~S.; Ghosh,~S.~K.; Strydom,~A.~M. Double-phase transition and giant positive magnetoresistance in the quasi-skutterudite Gd$_3$Ir$_4$Sn$_{13}$. \emph{J. Appl. Phys.} \textbf{2016}, \emph{119}, 123901\relax
\mciteBstWouldAddEndPuncttrue
\mciteSetBstMidEndSepPunct{\mcitedefaultmidpunct}
{\mcitedefaultendpunct}{\mcitedefaultseppunct}\relax
\EndOfBibitem
\bibitem[Remeika \latin{et~al.}(1980)Remeika, Espinosa, Cooper, Barz, Rowell, McWhan, Vandenberg, Moncton, Fisk, Woolf, Hamaker, Maple, Shirane, and Thomlinson]{Remeika1980}
Remeika,~J.~P.; Espinosa,~G.~P.; Cooper,~A.~S.; Barz,~H.; Rowell,~J.~M.; McWhan,~D.~B.; Vandenberg,~J.~M.; Moncton,~D.~E.; Fisk,~Z.; Woolf,~L.~D.; Hamaker,~H.~C.; Maple,~M.~B.; Shirane,~G.; Thomlinson,~W. A new family of ternary intermetallic superconducting/magnetic stannides. \emph{Solid State Commun.} \textbf{1980}, \emph{34}, 923--926\relax
\mciteBstWouldAddEndPuncttrue
\mciteSetBstMidEndSepPunct{\mcitedefaultmidpunct}
{\mcitedefaultendpunct}{\mcitedefaultseppunct}\relax
\EndOfBibitem
\bibitem[Oduchi \latin{et~al.}(2007)Oduchi, Tonohiro, Thamizhavel, Nakashima, Morimoto, Matsuda, Haga, Sugiyama, Takeuchi, Settai, Hagiwara, and Onuki]{Oduchi2007}
Oduchi,~Y.; Tonohiro,~C.; Thamizhavel,~A.; Nakashima,~H.; Morimoto,~S.; Matsuda,~T.~D.; Haga,~Y.; Sugiyama,~K.; Takeuchi,~T.; Settai,~R.; Hagiwara,~M.; Onuki,~Y. Magnetic properties of Ce$_3$T$_4$Sn$_{13}$ and Pr$_3$T$_4$Sn$_{13}$ (T = Co and Rh) single crystals. \emph{J. Magn. Magn. Mater} \textbf{2007}, \emph{310}, 249--251\relax
\mciteBstWouldAddEndPuncttrue
\mciteSetBstMidEndSepPunct{\mcitedefaultmidpunct}
{\mcitedefaultendpunct}{\mcitedefaultseppunct}\relax
\EndOfBibitem
\bibitem[Espinosa \latin{et~al.}(1982)Espinosa, Cooper, and Barz]{Espinosa1982}
Espinosa,~G.~P.; Cooper,~A.~S.; Barz,~H. Isomorphs of the superconducting/magnetic ternary stannides. \emph{Mater. Res. Bull.} \textbf{1982}, \emph{17}, 963--969\relax
\mciteBstWouldAddEndPuncttrue
\mciteSetBstMidEndSepPunct{\mcitedefaultmidpunct}
{\mcitedefaultendpunct}{\mcitedefaultseppunct}\relax
\EndOfBibitem
\bibitem[Hodeau \latin{et~al.}(1980)Hodeau, Chenavas, Marezio, and Remeika]{Hodeau1980}
Hodeau,~J.; Chenavas,~J.; Marezio,~M.; Remeika,~J. The crystal structure of SnYb$_3$Rh$_4$Sn$_{12}$, a new ternary superconducting stannide. \emph{Solid State Commun.} \textbf{1980}, \emph{36}, 839--845\relax
\mciteBstWouldAddEndPuncttrue
\mciteSetBstMidEndSepPunct{\mcitedefaultmidpunct}
{\mcitedefaultendpunct}{\mcitedefaultseppunct}\relax
\EndOfBibitem
\bibitem[Ślebarski and Goraus(2013)Ślebarski, and Goraus]{Ślebarski2013}
Ślebarski,~A.; Goraus,~J. Electronic structure and crystallographic properties of skutterudite-related Ce$_3$M$_4$Sn$_{13}$ and La$_3$M$_4$Sn$_{13}$ (M=Co, Ru, and Rh). \emph{Phys. Rev. B} \textbf{2013}, \emph{88}, 155122\relax
\mciteBstWouldAddEndPuncttrue
\mciteSetBstMidEndSepPunct{\mcitedefaultmidpunct}
{\mcitedefaultendpunct}{\mcitedefaultseppunct}\relax
\EndOfBibitem
\bibitem[Zhong \latin{et~al.}(2009)Zhong, Lei, and Mao]{Zhong2009}
Zhong,~G.; Lei,~X.; Mao,~J. Chemical bonding, electronic, and magnetic properties of R$_3$Co$_4$Sn$_{13}$ intermetallics (R=La, Ce, Sm, Gd, and Tb): Density functional calculations. \emph{Phys. Rev. B} \textbf{2009}, \emph{79}, 094424\relax
\mciteBstWouldAddEndPuncttrue
\mciteSetBstMidEndSepPunct{\mcitedefaultmidpunct}
{\mcitedefaultendpunct}{\mcitedefaultseppunct}\relax
\EndOfBibitem
\bibitem[Neha \latin{et~al.}(2016)Neha, Srivastava, Jha, Shruti, Awana, and Patnaik]{Neha2016}
Neha,~P.; Srivastava,~P.; Jha,~R.; Shruti; Awana,~V.~P.; Patnaik,~S. Improved superconducting properties of La$_3$Co$_4$Sn$_{13}$ with indium substitution. \emph{J. Alloys Compd.} \textbf{2016}, \emph{665}, 333--338\relax
\mciteBstWouldAddEndPuncttrue
\mciteSetBstMidEndSepPunct{\mcitedefaultmidpunct}
{\mcitedefaultendpunct}{\mcitedefaultseppunct}\relax
\EndOfBibitem
\bibitem[Scheie(2021)]{Scheie2021}
Scheie,~A. PyCrystalField: software for calculation, analysis and fitting of crystal electric field Hamiltonians. \emph{J. Appl. Cryst} \textbf{2021}, \emph{54}, 356--362\relax
\mciteBstWouldAddEndPuncttrue
\mciteSetBstMidEndSepPunct{\mcitedefaultmidpunct}
{\mcitedefaultendpunct}{\mcitedefaultseppunct}\relax
\EndOfBibitem
\bibitem[Lin \latin{et~al.}(2022)Lin, Li, Yu, Chen, Attfield, and Xing]{D1CS00563D}
Lin,~K.; Li,~Q.; Yu,~R.; Chen,~J.; Attfield,~J.~P.; Xing,~X. Chemical pressure in functional materials. \emph{Chem. Soc. Rev.} \textbf{2022}, \emph{51}, 5351--5364\relax
\mciteBstWouldAddEndPuncttrue
\mciteSetBstMidEndSepPunct{\mcitedefaultmidpunct}
{\mcitedefaultendpunct}{\mcitedefaultseppunct}\relax
\EndOfBibitem
\bibitem[Giannozzi \latin{et~al.}(2009)Giannozzi, Baroni, Bonini, Calandra, Car, Cavazzoni, Ceresoli, Chiarotti, Cococcioni, Dabo, Dal~Corso, de~Gironcoli, Fabris, Fratesi, Gebauer, Gerstmann, Gougoussis, Kokalj, Lazzeri, Martin-Samos, Marzari, Mauri, Mazzarello, Paolini, Pasquarello, Paulatto, Sbraccia, Scandolo, Sclauzero, Seitsonen, Smogunov, Umari, and Wentzcovitch]{Giannozzi_2009}
Giannozzi,~P. \latin{et~al.}  QUANTUM ESPRESSO: a modular and open-source software project for quantum simulations of materials. \emph{J. Phys. Condens. Matter} \textbf{2009}, \emph{21}, 395502\relax
\mciteBstWouldAddEndPuncttrue
\mciteSetBstMidEndSepPunct{\mcitedefaultmidpunct}
{\mcitedefaultendpunct}{\mcitedefaultseppunct}\relax
\EndOfBibitem
\bibitem[Perdew and Zunger(1981)Perdew, and Zunger]{perdew-zunger}
Perdew,~J.~P.; Zunger,~A. Self-interaction correction to density-functional approximations for many-electron systems. \emph{Phys. Rev. B} \textbf{1981}, \emph{23}, 5048--5079\relax
\mciteBstWouldAddEndPuncttrue
\mciteSetBstMidEndSepPunct{\mcitedefaultmidpunct}
{\mcitedefaultendpunct}{\mcitedefaultseppunct}\relax
\EndOfBibitem
\bibitem[Bellaiche and Vanderbilt(2000)Bellaiche, and Vanderbilt]{vca}
Bellaiche,~L.; Vanderbilt,~D. Virtual crystal approximation revisited: Application to dielectric and piezoelectric properties of perovskites. \emph{Phys. Rev. B} \textbf{2000}, \emph{61}, 7877--7882\relax
\mciteBstWouldAddEndPuncttrue
\mciteSetBstMidEndSepPunct{\mcitedefaultmidpunct}
{\mcitedefaultendpunct}{\mcitedefaultseppunct}\relax
\EndOfBibitem
\bibitem[Villars \latin{et~al.}(2006)Villars, Okamoto, and Cenzual]{villars2006asm}
Villars,~P.; Okamoto,~H.; Cenzual,~K. ASM alloy phase diagrams database. \emph{ASM International, Materials Park, OH, USA} \textbf{2006}, \relax
\mciteBstWouldAddEndPunctfalse
\mciteSetBstMidEndSepPunct{\mcitedefaultmidpunct}
{}{\mcitedefaultseppunct}\relax
\EndOfBibitem
\bibitem[\ifmmode~\acute{S}\else \'{S}\fi{}lebarski \latin{et~al.}(2015)\ifmmode~\acute{S}\else \'{S}\fi{}lebarski, Goraus, Witas, Kalinowski, and Fija\l{}kowski]{Slebarski2015}
\ifmmode~\acute{S}\else \'{S}\fi{}lebarski,~A.; Goraus,~J.; Witas,~P.; Kalinowski,~L.; Fija\l{}kowski,~M. Study of $d$-electron correlations in skutterudite-related Ce$_3$M$_4$Sn$_{13}$ (M = Co, Ru, and Rh). \emph{Phys. Rev. B} \textbf{2015}, \emph{91}, 035101\relax
\mciteBstWouldAddEndPuncttrue
\mciteSetBstMidEndSepPunct{\mcitedefaultmidpunct}
{\mcitedefaultendpunct}{\mcitedefaultseppunct}\relax
\EndOfBibitem
\bibitem[Mishra \latin{et~al.}(2011)Mishra, Schwickert, Langer, and P$\ddot{o}$ttgen]{Mishra2011}
Mishra; Schwickert,~C.; Langer,~T.; P$\ddot{o}$ttgen,~R. Ternary Stannides RE$_3$Ru$_4$Sn$_{13}$ (RE = La, Ce, Pr, Nd) – Structure, Magnetic Properties, and $^{119}$Sn M$\ddot{o}$ssbauer Spectroscopy. \emph{Z. f$\ddot{u}$r Naturforsch. - B} \textbf{2011}, \emph{66}, 664--670\relax
\mciteBstWouldAddEndPuncttrue
\mciteSetBstMidEndSepPunct{\mcitedefaultmidpunct}
{\mcitedefaultendpunct}{\mcitedefaultseppunct}\relax
\EndOfBibitem
\bibitem[\ifmmode~\acute{S}\else \'{S}\fi{}lebarski \latin{et~al.}(2013)\ifmmode~\acute{S}\else \'{S}\fi{}lebarski, Fija\l{}kowski, and Goraus]{Slebarski2013}
\ifmmode~\acute{S}\else \'{S}\fi{}lebarski,~A.; Fija\l{}kowski,~M.; Goraus,~J. Evolution from a magnetically correlated state to a single impurity state in heavy fermion system Ce$_{3-x}$La$_x$Co$_4$Sn$_{13}$. \emph{phys. status solidi (b)} \textbf{2013}, \emph{250}, 472--475\relax
\mciteBstWouldAddEndPuncttrue
\mciteSetBstMidEndSepPunct{\mcitedefaultmidpunct}
{\mcitedefaultendpunct}{\mcitedefaultseppunct}\relax
\EndOfBibitem
\bibitem[Ranjith \latin{et~al.}(2015)Ranjith, Nath, Skoulatos, Keller, Kasinathan, Skourski, and Tsirlin]{ranjith2015collinear}
Ranjith,~K.~M.; Nath,~R.; Skoulatos,~M.; Keller,~L.; Kasinathan,~D.; Skourski,~Y.; Tsirlin,~A.~A. Collinear order in the frustrated three-dimensional spin-1/2 antiferromagnet Li$_2$CuW$_2$O$_8$. \emph{Phys. Rev. B} \textbf{2015}, \emph{92}, 094426\relax
\mciteBstWouldAddEndPuncttrue
\mciteSetBstMidEndSepPunct{\mcitedefaultmidpunct}
{\mcitedefaultendpunct}{\mcitedefaultseppunct}\relax
\EndOfBibitem
\bibitem[Godart \latin{et~al.}(1993)Godart, Mazumdar, Dhar, Nagarajan, Gupta, Padalia, and Vijayaraghavan]{Godart1993}
Godart,~C.; Mazumdar,~C.; Dhar,~S.~K.; Nagarajan,~R.; Gupta,~L.~C.; Padalia,~B.~D.; Vijayaraghavan,~R. Valence state of Sm in SmRuSn$_3$. \emph{Phys. Rev. B} \textbf{1993}, \emph{48}, 16402--16406\relax
\mciteBstWouldAddEndPuncttrue
\mciteSetBstMidEndSepPunct{\mcitedefaultmidpunct}
{\mcitedefaultendpunct}{\mcitedefaultseppunct}\relax
\EndOfBibitem
\bibitem[Fukuhara \latin{et~al.}(1991)Fukuhara, Sakamoto, and Sato]{Fukuhara_1991}
Fukuhara,~T.; Sakamoto,~I.; Sato,~H. Transport and magnetic properties of RERuSn$_3$ (RE=La, Ce, Pr, Nd, Sm): a heavy fermion compound CeRuSn$_3$ and a new valence fluctuating compound SmRuSn$_3$. \emph{J Phys. Condens. Matter} \textbf{1991}, \emph{3}, 8917\relax
\mciteBstWouldAddEndPuncttrue
\mciteSetBstMidEndSepPunct{\mcitedefaultmidpunct}
{\mcitedefaultendpunct}{\mcitedefaultseppunct}\relax
\EndOfBibitem
\bibitem[Nair \latin{et~al.}(2016)Nair, K., Britz, Ghosh, Reinke, and Strydom]{NAIR2016254}
Nair,~H.~S.; K.,~R.~K.; Britz,~D.; Ghosh,~S.~K.; Reinke,~C.; Strydom,~A.~M. Field-insensitive heavy fermion features and phase transition in the caged-structure quasi-skutterudite Sm$_3$Ru$_4$Ge$_{3}$. \emph{J. Alloys Compd.} \textbf{2016}, \emph{669}, 254--261\relax
\mciteBstWouldAddEndPuncttrue
\mciteSetBstMidEndSepPunct{\mcitedefaultmidpunct}
{\mcitedefaultendpunct}{\mcitedefaultseppunct}\relax
\EndOfBibitem
\bibitem[Mazumdar \latin{et~al.}(1996)Mazumdar, Hossain, Nagarajan, Godart, Dhar, Gupta, Padalia, and Vijayaraghavan]{Mazumdar1996TransportSmRuSn_3}
Mazumdar,~C.; Hossain,~Z.; Nagarajan,~R.; Godart,~C.; Dhar,~S.~K.; Gupta,~L.~C.; Padalia,~B.~D.; Vijayaraghavan,~R. Transport, magnetic, and $^{119}$Sn M\"{o}ssbauer studies on magnetically ordered valence fluctuating compound SmRuSn$_3$. \emph{J Appl. Phys.} \textbf{1996}, \emph{79}, 6349--6351\relax
\mciteBstWouldAddEndPuncttrue
\mciteSetBstMidEndSepPunct{\mcitedefaultmidpunct}
{\mcitedefaultendpunct}{\mcitedefaultseppunct}\relax
\EndOfBibitem
\bibitem[Mugiraneza and Hallas(2022)Mugiraneza, and Hallas]{Mugiraneza2022}
Mugiraneza,~S.; Hallas,~A.~M. Tutorial: a beginner’s guide to interpreting magnetic susceptibility data with the Curie-Weiss law. \emph{Commun. Phys.} \textbf{2022}, \emph{5}, 95\relax
\mciteBstWouldAddEndPuncttrue
\mciteSetBstMidEndSepPunct{\mcitedefaultmidpunct}
{\mcitedefaultendpunct}{\mcitedefaultseppunct}\relax
\EndOfBibitem
\bibitem[Peratheepan and Strydom(2015)Peratheepan, and Strydom]{Peratheepan_2015}
Peratheepan,~P.; Strydom,~A.~M. Electronic, magnetic, and transport properties of the isotypic aluminides SmT$_2$Al$_{10}$ (T = Fe, Ru). \emph{J. Phys. Condens. Matter} \textbf{2015}, \emph{27}, 095604\relax
\mciteBstWouldAddEndPuncttrue
\mciteSetBstMidEndSepPunct{\mcitedefaultmidpunct}
{\mcitedefaultendpunct}{\mcitedefaultseppunct}\relax
\EndOfBibitem
\bibitem[Khoury \latin{et~al.}(2022)Khoury, Han, Jovanovic, Singha, Song, Queiroz, Ong, and Schoop]{khoury2022class}
Khoury,~J.~F.; Han,~B.; Jovanovic,~M.; Singha,~R.; Song,~X.; Queiroz,~R.; Ong,~N.-P.; Schoop,~L.~M. A Class of Magnetic Topological Material Candidates with Hypervalent Bi Chains. \emph{J. Am. Chem. Soc.} \textbf{2022}, \emph{144}, 9785--9796\relax
\mciteBstWouldAddEndPuncttrue
\mciteSetBstMidEndSepPunct{\mcitedefaultmidpunct}
{\mcitedefaultendpunct}{\mcitedefaultseppunct}\relax
\EndOfBibitem
\bibitem[Higashinaka \latin{et~al.}(2011)Higashinaka, Maruyama, Nakama, Miyazaki, Aoki, and Sato]{Higashinaka2011}
Higashinaka,~R.; Maruyama,~T.; Nakama,~A.; Miyazaki,~R.; Aoki,~Y.; Sato,~H. Unusual Field-Insensitive Phase Transition and Kondo Behavior in SmTi$_2$Al$_{20}$. \emph{J. Phys. Soc. Jpn.} \textbf{2011}, \emph{80}, 093703\relax
\mciteBstWouldAddEndPuncttrue
\mciteSetBstMidEndSepPunct{\mcitedefaultmidpunct}
{\mcitedefaultendpunct}{\mcitedefaultseppunct}\relax
\EndOfBibitem
\bibitem[Sanada \latin{et~al.}(2005)Sanada, Aoki, Aoki, Tsuchiya, Kikuchi, Sugawara, and Sato]{sanada2005exotic}
Sanada,~S.; Aoki,~Y.; Aoki,~H.; Tsuchiya,~A.; Kikuchi,~D.; Sugawara,~H.; Sato,~H. Exotic heavy-fermion state in filled skutterudite SmOs$_4$Sb$_{12}$. \emph{J. Phys. Soc. Jpn.} \textbf{2005}, \emph{74}, 246--249\relax
\mciteBstWouldAddEndPuncttrue
\mciteSetBstMidEndSepPunct{\mcitedefaultmidpunct}
{\mcitedefaultendpunct}{\mcitedefaultseppunct}\relax
\EndOfBibitem
\bibitem[Denton and Ashcroft(1991)Denton, and Ashcroft]{denton1991vegard}
Denton,~A.~R.; Ashcroft,~N.~W. Vegard’s law. \emph{Phys. Rev. A} \textbf{1991}, \emph{43}, 3161\relax
\mciteBstWouldAddEndPuncttrue
\mciteSetBstMidEndSepPunct{\mcitedefaultmidpunct}
{\mcitedefaultendpunct}{\mcitedefaultseppunct}\relax
\EndOfBibitem
\end{mcitethebibliography}

\providecommand{\latin}[1]{#1}
\makeatletter
\providecommand{\doi}
  {\begingroup\let\do\@makeother\dospecials
  \catcode`\{=1 \catcode`\}=2 \doi@aux}
\providecommand{\doi@aux}[1]{\endgroup\texttt{#1}}
\makeatother
\providecommand*\mcitethebibliography{\thebibliography}
\csname @ifundefined\endcsname{endmcitethebibliography}  {\let\endmcitethebibliography\endthebibliography}{}

For Table of Contents Only
\begin{figure}[H]
\centering
 \includegraphics[width=\textwidth]{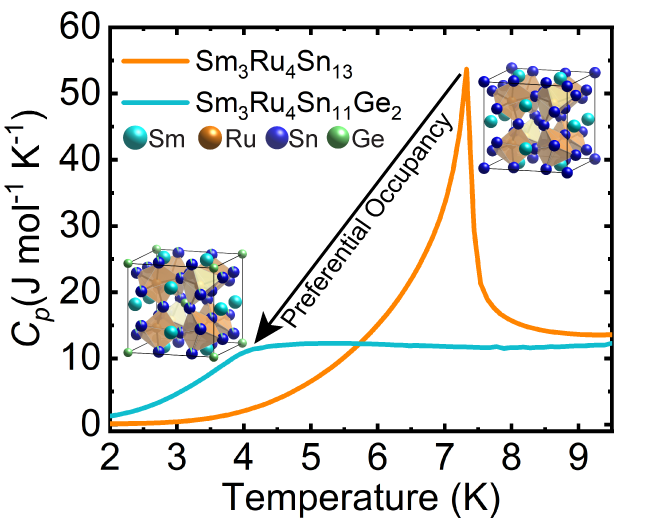}
\end{figure}

\end{document}


\newpage
\tableofcontents
\newpage

\section{PyCrystalField}
PyCrystalField was utilized to calculate the eigenvalues of degenerate orbitals within Sm$^{3+}$.$^1$ This method calculates the eigenvalues using a point-charge model from the CIFs obtained from single-crystal XRD. 

\begin{table}[hbt!]
    \centering
    \begin{tabular}{|c|c|c|c|}
        \hline
         &  \textbf{Sm$_3$Ru$_4$Sn$_{13}$}&  \textbf{Sm$_3$Ru$_4$Sn$_{12}$Ge}& \textbf{Sm$_3$Ru$_4$Sn$_{11}$Ge$_2$}\\
        \hline
        \multirow[c]{6}{*}{\textbf{E (meV)}} & 0.000 &  0.000 & 0.000 \\
         & 0.000 &  0.000 & 0.000 \\
         &  1.875&  11.580& 25.163\\
         &  1.875&  11.580& 25.163\\
         &  3.172&  38.206& 39.021\\
         &  3.172&  38.206& 39.021\\
        \hline
    \end{tabular}
    \caption{Eigenvalues of the degenerate orbitals in Sm$^{3+}$ in a D$_{2d}$ site-symmetry.}
    \label{tab:Eigenvalues}
\end{table}

\section{Energy Dispersive Spectroscopy (EDS)}
Energy dispersive spectroscopy analysis was performed using a Zeiss Auriga FIB/SEM with an Oxford X-Max detector. Map analysis was collected with an aperture size of 60.00 $\mu$m at 20.000 kV on flux-grown crystals of Sm$_3$Ru$_4$Sn$_{13-x}$Ge$_x$.
\begin{figure}[hbt!]
  \includegraphics[width=\textwidth]{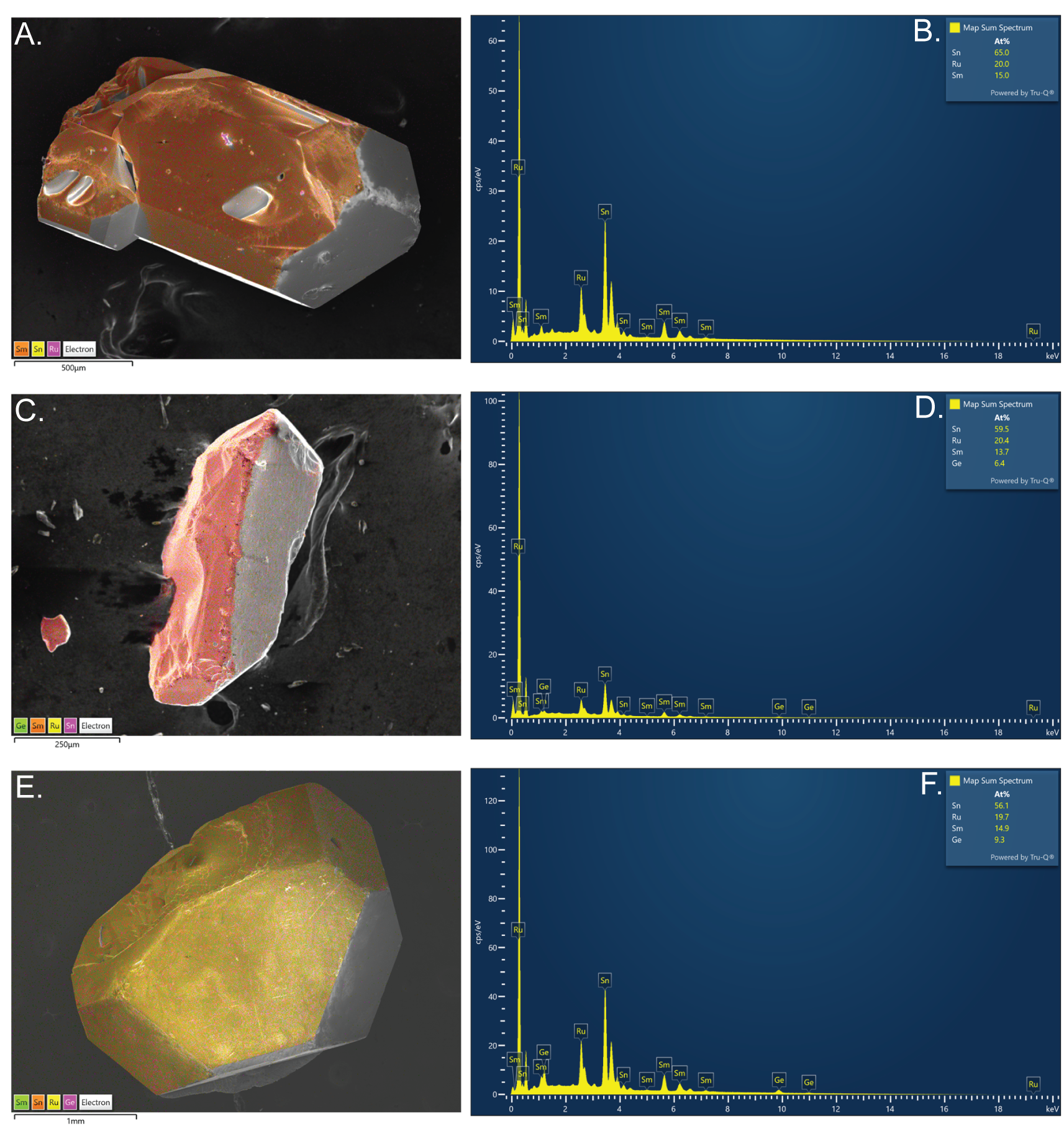}
  \caption{EDS images of A) Sm$_3$Ru$_4$Sn$_{13}$ and its compositional analysis (B), C) Sm$_3$Ru$_4$Sn$_{12}$Ge and its compositional analysis (D), and E) Sm$_3$Ru$_4$Sn$_{11}$Ge$_2$ and its compositional analysis (F). The composition of Sm$_3$Ru$_4$Sn$_{13}$ perfectly matches the 15:20:65 ratio, while Sm$_3$Ru$_4$Sn$_{12}$Ge and Sm$_3$Ru$_4$Sn$_{11}$Ge$_2$ are slightly off from the ideal ratio.}
  \label{SI EDS}
  \centering
  \end{figure}
  
\section{Single-Crystal X-Ray Diffraction}
\begin{table}[hbt!]
    \centering
    \begin{tabular}{|c|c|c|c|c|c|c|}
        \hline
        \textbf{Label} & \textbf{Wyck.} & \textbf{x} & \textbf{y} & \textbf{z} & \textbf{Occupancy} & \textbf{U$_{eq}$$^*$} \\
        \hline
        \multicolumn{7}{|c|}{\textbf{Sm$_3$Ru$_4$Sn$_{13}$}} \\
        \hline
        Sm    &6\textit{c}&3/4&0&1/2&1&5(1)\\
        Sn(1) &2\textit{a}&1/2&1/2&1/2&1&8(1)\\
        Sn(2) &24\textit{k}&0.8047(1)&0.3457(1)&1/2&1&6(1)\\
        Ru    &8\textit{e}&3/4&1/4&1/4&1&4(1)\\
        \hline
        \multicolumn{7}{|c|}{\textbf{Sm$_3$Ru$_4$Sn$_{12}$Ge}} \\
        \hline
        Sm    &6\textit{c}&1/2&1&1/4&1&6(1)\\
        Ge    &2\textit{a}&1/2&1/2&1/2&0.60(3)&10(1)\\
        Sn(1) &2\textit{a}&1/2&1/2&1/2&0.40(3)&10(1)\\
        Sn(2) &24\textit{k}&1/2&0.6547(1)&0.1940(1)&0.988(4)&8(1)\\
        Ru    &8\textit{e}&3/4&3/4&1/4&1&6(1)\\
        \hline
        \multicolumn{7}{|c|}{\textbf{Sm$_3$Ru$_4$Sn$_{11}$Ge$_2$}} \\
        \hline
        Sm    &6\textit{c}&0&1/2&3/4&1&7(1)\\
        Ge(1) &2\textit{a}&1/2&1/2&1/2&1&12(1)\\
        Ru    &8\textit{e}&1/4&1/4&3/4&1&8(1)\\
        Sn    &24\textit{k}&0.3451(1)&1/2&0.8066(1)&0.937(19)&10(1)\\
        Ge(2) &24\textit{k}&0.3451(1)&1/2&0.8066(1)&0.063(19)&10(1)\\
        \hline
        \multicolumn{7}{|c|}{$^*$U$_{eq}$ is defined as one third of the trace of the orthogonalized U$_{ij}$ tensor.} \\
        \hline
    \end{tabular}
    \caption{Atomic coordinates ($\times$10$^4$) and equivalent isotropic displacement parameters (\AA$^2$$\times$10$^3$) for  Sm$_3$Ru$_4$Sn$_{13-x}$Ge$_x$ (x = 0, 1, or 2) at 100.00 K with estimated standard deviations in parentheses.}
    \label{tab:SI Atom Coord}
\end{table}

\section{Powder X-Ray Diffraction}
Powder x-ray diffraction data was collected by grinding $\sim$ 10mg  of Sm$_{3}$Ru$_{4}$Sn$_{13-x}$Ge$_{x}$ into a fine powder using an agate mortar and pestle. The powder was placed into a Bruker D6 Phaser diffractometer using Cu K$\alpha$ radiation ($\lambda$ = 1.5406$\mathring{A}$) with a Ni K$\beta$ filter. Rietveld refinement was performed using TOPAS software.

\begin{figure}[hbt!]
  \includegraphics[width=\textwidth]{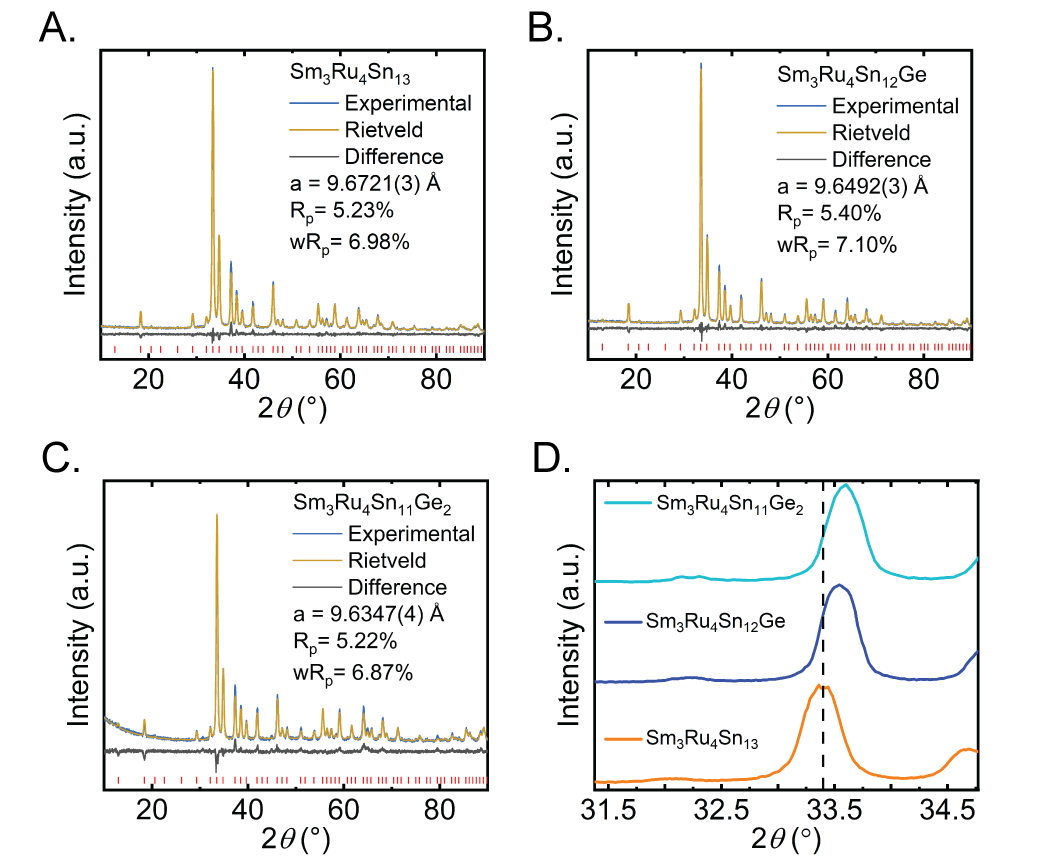}
  \caption{Rietveld refinement of A) Sm$_3$Ru$_4$Sn$_{13}$, B) Sm$_3$Ru$_4$Sn$_{12}$Ge, and C) Sm$_3$Ru$_4$Sn$_{11}$Ge$_2$ fit to \textit{Pm$\bar{3}$n} space group. The unit cell parameter decreases as Ge concentration increases. D) The main peak at 33.4$^{\circ}$ shifts to higher angles as the concentration of Ge increases.}
  \label{SI PXRD}
  \centering
\end{figure}

\section{X-ray Photoelectron Spectroscopy}
\begin{figure}[hbt!]
  \includegraphics[width=\textwidth]{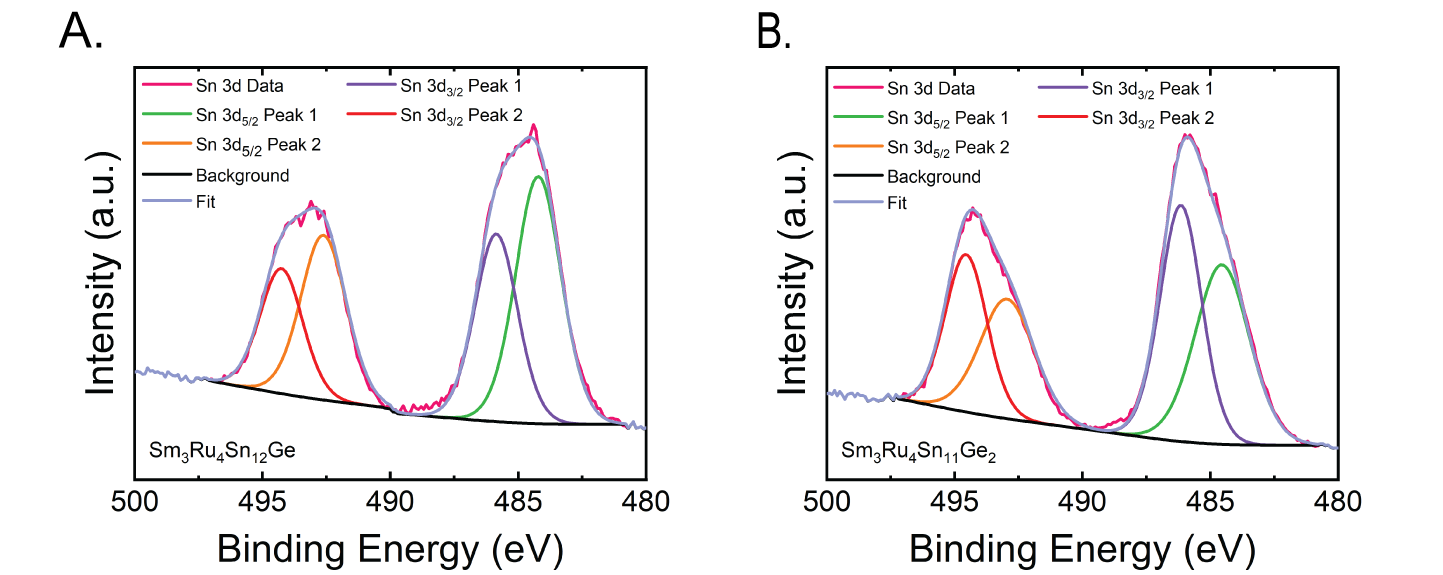}
  \caption{XPS data of A) Sm$_3$Ru$_4$Sn$_{12}$Ge and B) Sm$_3$Ru$_4$Sn$_{11}$Ge$_2$. As the concentration of Ge increases, Sn 3\textit{d}$_{5/2}$ starts to decrease as  to Sn 3\textit{d}$_{3/2}$ increases. The shift to higher binding energies is due to stronger covalent bonding between Ru--Sn(24\textit{k})}
  \label{SI XPS}
  \centering
\end{figure}

\section{Magnetometry}

\begin{figure}[hbt!]
  \includegraphics[width=0.55\textwidth]{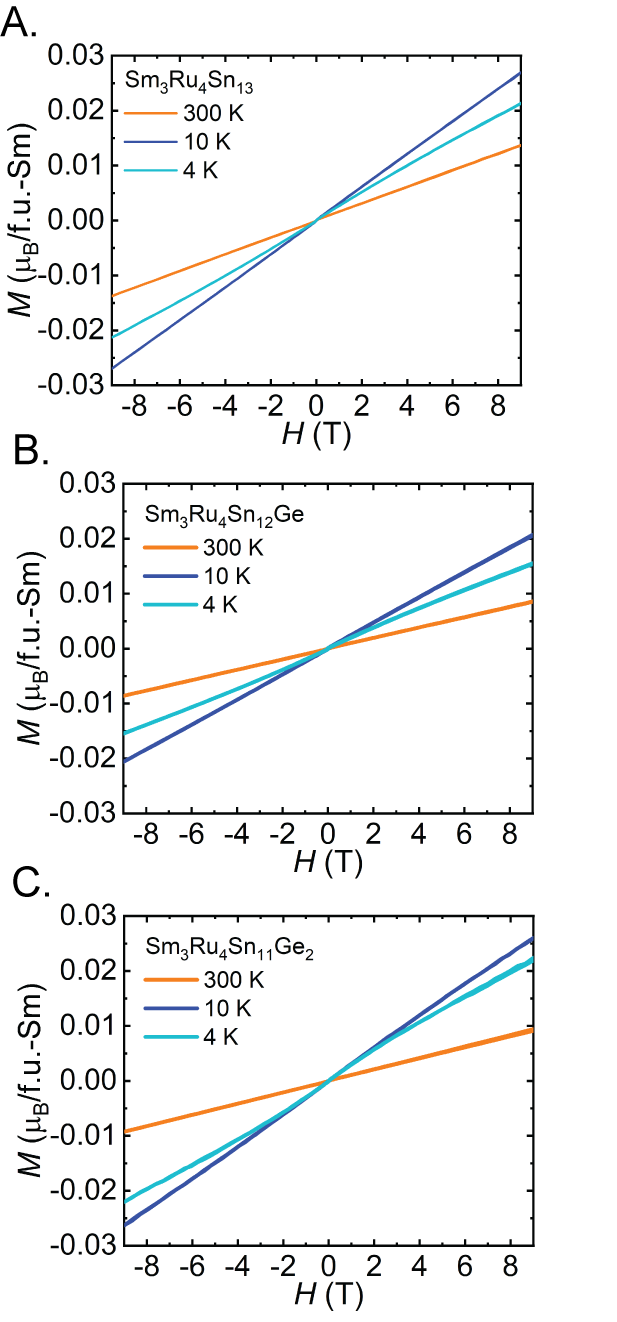}
  \caption{Magnetic moment versus applied magnetic field of  A) Sm$_3$Ru$_4$Sn$_{13}$, B) Sm$_3$Ru$_4$Sn$_{12}$Ge, and C) Sm$_3$Ru$_4$Sn$_{11}$Ge$_2$ from 0 T to 9 T at 300 K, 10 K, and 4 K. None of the compounds achieved saturation up to 9 T.}
  \label{SI MH}
  \centering
\end{figure}

\begin{figure}[hbt!]
  \includegraphics[width=\textwidth]{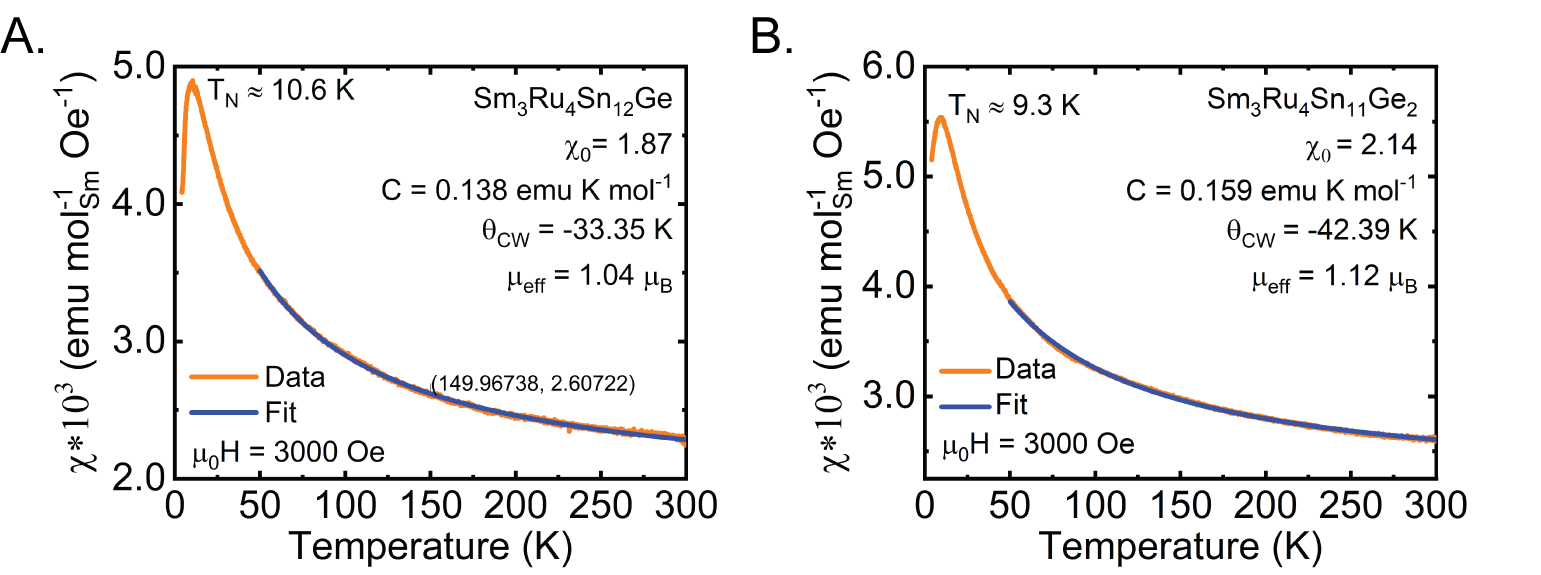}
  \caption{The field-cooled (FC) magnetic susceptibility of A) Sm$_3$Ru$_4$Sn$_{12}$Ge and B) Sm$_3$Ru$_4$Sn$_{11}$Ge$_2$ from 2--200 K. A broad antiferromagnetic transition is present at 10.6 K and 9.5 K for x = 1 and 2, respectively. The fitted Curie-Weiss data gave $\theta$$_{CW}$ = -33.35 K and -42.39 K for x = 1 and 2, respectively. The $\mu$$_{eff}$ for x = 1 and 2 is 1.04 and 1.12 $\mu$$_{B}$, respectively, which is slightly higher than the theoretical $\mu$$_{eff}$ of 0.85 for Sm$^{3+}$.}
  \label{fig:SI MT}
  \centering
\end{figure}

\section{Heat Capacity}
\begin{figure}[hbt!]
  \includegraphics[width=\textwidth]{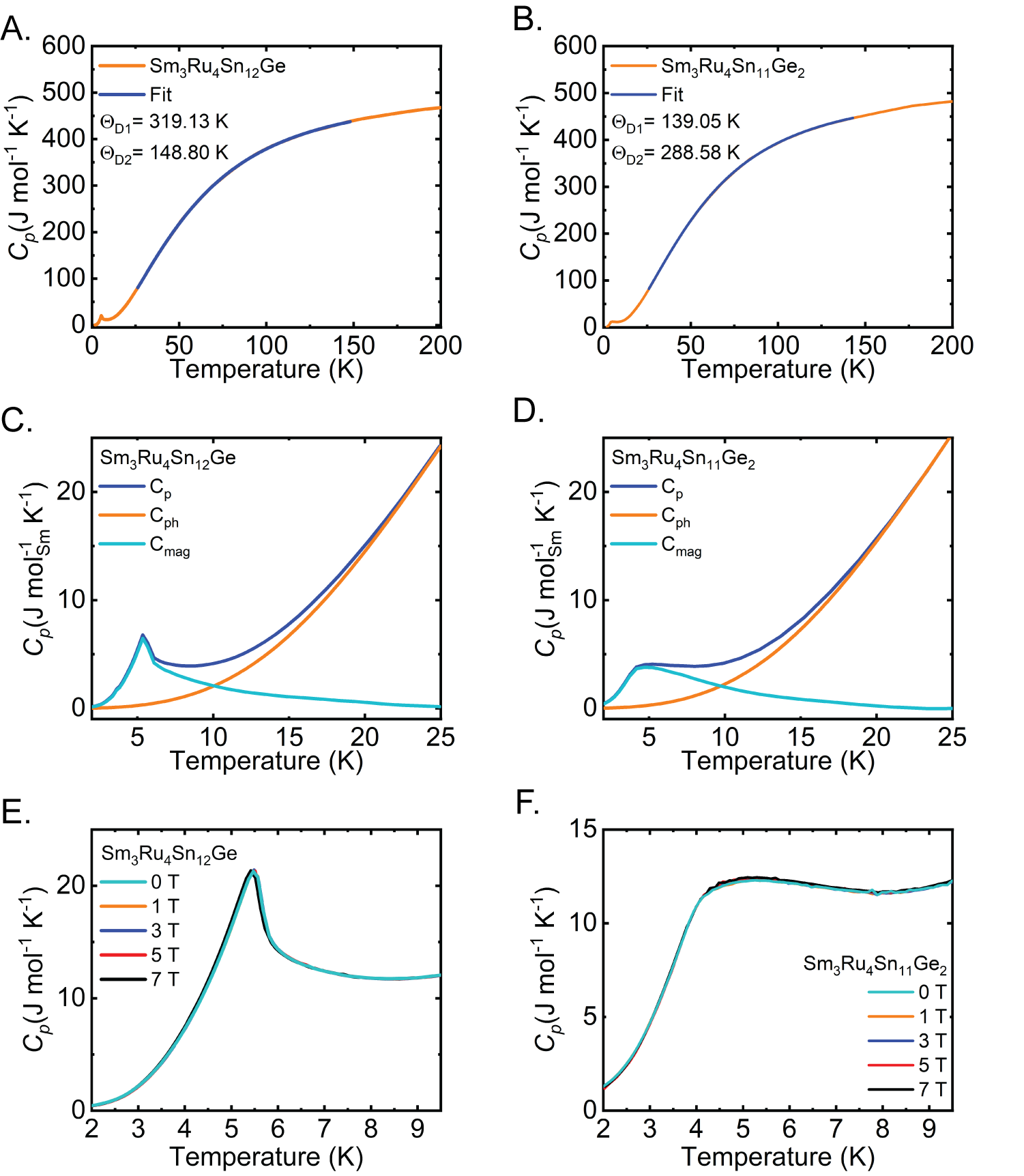}
  \caption{The two-mode Debye fit between 25 K to 150 K for A) Sm$_3$Ru$_4$Sn$_{12}$Ge and B) Sm$_3$Ru$_4$Sn$_{11}$Ge$_2$. The Debye temperatures are 319.13 K and 148.80 K for Sm$_3$Ru$_4$Sn$_{12}$Ge and 139.05 K and 288.58 K for Sm$_3$Ru$_4$Sn$_{11}$Ge$_2$. The sum of the oscillator terms was $\sim$20.1 and 20.5 for x = 1 and 2, respectively, in good agreement with their stoichiometries. The fittings for C$_p$and C$_{ph}$ from the two-mode Debye fits are shown for C) Sm$_3$Ru$_4$Sn$_{12}$Ge and D) Sm$_3$Ru$_4$Sn$_{11}$Ge$_2$. E) Sm$_3$Ru$_4$Sn$_{12}$Ge and F) Sm$_3$Ru$_4$Sn$_{11}$Ge$_2$ show field-insensitive behavior.}  
  \label{SI Cp}
  \centering
  \end{figure}
\section{References}